\documentclass[preprint]{aastex}
\usepackage[section] {placeins}
\usepackage[T1]{fontenc}
\usepackage[usenames]{xcolor}

\shorttitle{Alfv\'{e}n Ionization}
\shortauthors{Stark et al.}

\begin{document}

\title{Ionization in Atmospheres of Brown Dwarfs and Extrasolar Planets V:  \\Alfv\'{e}n Ionization\footnote{Paper accepted to The Astrophysical Journal}}

\author{C. R. Stark\altaffilmark{1}, Ch. Helling\altaffilmark{1}, D. A. Diver\altaffilmark{2} and P. B. Rimmer\altaffilmark{1} }
\altaffiltext{1}{SUPA, School of Physics and Astronomy, University of St Andrews, St Andrews, KY16 9SS, Scotland, UK.}
\altaffiltext{2}{SUPA, School of Physics and Astronomy, Kelvin Building, University of Glasgow, Glasgow, G12 8QQ, Scotland, UK.}

\email{craig.stark@st-andrews.ac.uk}

\begin{abstract}
Observations of continuous radio and sporadic X-ray emission from low-mass objects suggest they harbour localized plasmas in their atmospheric environments.  For low-mass objects, the degree of thermal ionization is insufficient to qualify the ionized component as a plasma, posing the question:  what ionization processes can efficiently produce the required plasma that is the source of the radiation?  We propose Alfv\'{e}n ionization as a mechanism for producing localized pockets of ionized gas in the atmosphere, having sufficient degrees of ionization ($\geq10^{-7}$) that they constitute plasmas.  We outline the criteria required for Alfv\'{e}n ionization and demonstrate it's applicability in the atmospheres of low-mass objects such as giant gas planets, brown dwarfs and M-dwarfs for both solar and sub-solar metallicities.  We find that Alfv\'{e}n ionization is most efficient at mid to low atmospheric pressures where a seed plasma is easier to magnetize and the pressure gradients needed to drive the required neutral flows are the smallest.  For the model atmospheres considered, our results show that degrees of ionization of $10^{-6}-1$ can be obtained as a result of Alfv\'{e}n ionization.  Observable consequences include continuum Bremsstrahlung emission, superimposed with spectral lines from the plasma ion species (e.g. He, Mg, H$_{2}$ or CO lines).  Forbidden lines are also expected from the metastable population.  The presence of an atmospheric plasma opens the door to a multitude of plasma and chemical processes not yet considered in current atmospheric models.  The occurrence of Alfv\'{e}n ionization may also be applicable to other astrophysical environments such as protoplanetary disks.
\end{abstract}


\section{Introduction}

Current atmospheric models of low-mass objects do not fully consider coherent plasma effects or collective plasma processes and how they fit into a holistic understanding of atmospheric phenomena such as cloud formation and evolution.  Observations of continuous radio and sporadic X-ray emission from low-mass objects suggest that such objects harbour an atmospheric, magnetized plasma of some form.  This suggests that atmospheres at the very least, are composed of gas-plasma mixtures where dust (and by extension clouds) can grow, creating a dusty plasma (a plasma containing charged particulates of nanometer and micrometer size).  Charged dust particles will be susceptible to inter-grain electrical discharges~\citep{helling2011a,helling2013} that will energize the ambient gas-plasma~\citep{craig2009,craig2013,li2009} and will participate in other collective plasma phenomena such as waves, flows and instabilities.  Therefore, it is key to our understanding of substellar atmospheres that the relevant plasma phenomena are incorporated into atmospheric models and their observational consequences quantified.  Before all else, it is critical to identify the ionization processes that create and replenish the atmospheric plasma, that may feed the extended envelope surrounding a low-mass object.

Low-mass objects with spectral type later than M7 (ultracool dwarfs) can be strong sources of radio emission which infers the presence of atmospheric plasmas.  Observations of the M9 dwarf  TVLM 513-46546 at 4.88 and 8.44~GHz characterize the radio emission as variable and with a periodicity consistent with the estimated period of rotation $\approx2$~hrs~\citep{hallinan2006}.  Follow-up observations identified additional periodic (1.96~hrs) bursts of 100\% circularly polarized, coherent radio emission~\citep{hallinan2007}.  Similar radio emission signatures have been detected from other ultracool dwarfs such as T6.5 dwarf 2MASS J1047+21~\citep{route2012}, the M8.5 dwarf LSR J1835+3259 and the L3.5 dwarf 2MASS J00361617+1821104, with brightness temperatures indicating that emission must come from a coherent source, most likely an electron cyclotron maser instability~\citep{hallinan2008}.   

The coherent nature of the emission allows the determination of the magnetic field strength; if an electron cyclotron maser is assumed the radiation is predominately emitted at the harmonics of the electron cyclotron frequency.  Modelling of the electron cyclotron maser emission mechanism has uncovered detail of the expected electromagnetic signature arising when the maser is electron-beam-driven, loss-cone-driven~\citep{yu2012} and driven by electrical currents resulting from an angular velocity shear~\citep{nichols2012}.  Assuming an electron cyclotron maser emission mechanism, the required magnetic flux densities of the ultracool dwarfs must be $\approx O(1~\textnormal{kG})$, which is consistent with other observed radio emitting ultracool dwarfs~\citep{mclean2011,mclean2012,osten2009,antonova2007, antonova2008,antonova2013,berger2006,berger2009}.  

Measurements of M-dwarf magnetic flux densities from the Zeeman effect, as a function of spectral type, show that $B\approx O(1~\textnormal{kG})$~\citep{shulyak2011,reiners2012}.  No direct observations have been made of exoplanetary magnetic fields but magnetic dynamo simulations suggest that magnetic flux densities of $\approx  O(10~\textnormal{G})$ can be expected~\citep{sanchez2004}.  This is consistent with predicted magnetic field strengths of massive exoplanets by~\cite{christensen2009}.  They suggest a scaling law for inferring the magnetic field strength based on energy-flux considerations, consistent with the maximum magnetic fields found in rapidly rotating low-mass stars and planetary magnetic fields such as Jupiter.  Transit observations of explanets may provide a method of deducing the magnitude of the planetary magnetic field by investigating the interaction between the stellar wind and the planetary magnetosphere~\citep{vidotto2011,vidotto2011b}.

Radio and X-ray observations suggest that plasmas are a potentially important component in the atmospheric environments of low-mass objects.  A plasma is defined as a collection of charged particles, of sufficiently high number density that the Coulomb force is significant in determining its properties, yet sufficiently dilute that the nearest neighbour  interaction is dominated not by binary collisions, but instead by the collective electromagnetic influence of the many distant particles.  Locally, the entirety of the gas does not need to be ionized for it to behave as a plasma, in fact for an ionized gas to be considered a plasma it only needs to be weakly ionized (i.e. partially ionized), provided the collective electromagnetic influence of the many distant particles is significant in determining the dynamics.  A measure of the extent  an ionized gas behaves like a plasma is the degree of ionization.  It is defined as the ratio of the density of the charged species $n_{i}$ to that of the neutral species $n_{gas}$, where $f_{e}=n_{i}/(n_{i}+n_{gas})=1$ is a completely ionized plasma and $f_{e}\approx10^{-7}$ is a weakly ionized plasma~\citep{diver2001,fridman2008}.  If localized pockets of plasma can be created and sustained in the atmosphere, this would open the door to a multitude of plasma processes and chemistry not yet considered in current models of substellar and planetary objects.  

Helling and co-workers have made progress in considering the effects of plasma phenomena in substellar atmospheres~\citep{helling2011a,helling2011b,helling2013}.~~\cite{helling2011b} used \textsc{Drift-Phoenix} model atmospheres to investigate the effect of dust-induced collisional ionization.  They found that ionization by turbulence-induced dust-dust collisions was the most efficient of the ionization processes considered but the electron density produced was insufficient to significantly improve the degree of ionization.  However, the resulting charged dust grains that compose the atmospheric clouds, are susceptible to inter-grain electrical discharge events~\citep{helling2011a}.  

In discharge events the electric field permeating the intervening ambient gas exceeds a threshold value resulting in the electrical breakdown of the gas.~~\cite{helling2013} investigated the electrical breakdown conditions under which mineral clouds in substellar atmospheres undergo discharge events.  Furthermore, the impact of these events on the local gas-phase chemistry, the effective temperature and the primordial gas-phase metallicity was assessed.  Needle-shaped grains can play an important role: the strong electric fields at the polar regions of the grain~\citep{stark2006} are particularly sensitive to inter-grain discharges~\citep{craig2009,craig2013,li2009} enhancing the effect.  An ensemble of such discharges, in analogy to laboratory microdischarges~\citep{becker2006}, will amplify the local degree of ionization to levels that the ionized gas constitutes a plasma~\citep{helling2011a}.  If the ambient magnetic field permeating the atmosphere is sufficiently strong the plasma will be magnetized, coupling the atmosphere to the magnetic field.

The question remains what ionization process can efficiently produce a plasma in the atmospheric environments of low-mass objects.  The degree of thermal ionization is insufficient to qualify the ionized component as a plasma; however, inter-grain electrical discharge events~\citep{helling2013} and cosmic-ray ionization processes~\citep{rimmer2013} can occur that enhance the local degree of ionization.  In this paper, we propose Alfv\'{e}n ionization (also known as {\it critical ionization velocity phenomenon}) as a simple mechanism for producing localized pockets of plasma in the atmospheres of low-mass objects.  In Alfv\'{e}n ionization a constant stream of neutral gas impinges on a low-density magnetized plasma.  The inflowing neutral atoms collide with and displace the plasma ions, leaving behind a significant charge imbalance that accelerates electrons to energies sufficient to ionize the local gas via electron-neutral impact ionization.  Alfv\'{e}n ionization requires an initial, low-density magnetized seed plasma and a neutral gas flow that reaches a critical threshold speed.  As well as creating an ionized plasma, Alfv\'{e}n ionization has significant implications for atmospheric chemistry.
 
The paper is structured as follows: Section~\ref{drift} summarises the non-phase-equilibrium atmosphere model (\textsc{Drift-Phoenix}) of Helling and co-workers of which a selected number of models will be used in the subsequent analysis;  Section~\ref{section2} introduces Alfv\'{e}n ionization and discusses it in the context of planetary, brown dwarf and M-dwarf atmospheres, defining the criteria of its applicability.  The emission levels of ultracool dwarfs are similar to that of early type active M-dwarfs~\citep{antonova2013} and so we are interested in the link connecting M-dwarfs, brown dwarfs and giant gas planets.  It is plausible that similar mechanisms for the production of the source plasma and the radio emission are at work.  Section~\ref{degree} quantifies the resulting degree of ionization expected from Alfv\'{e}n ionization;  and Section~\ref{section4} summarises our findings and discusses the consequent scientific implications.  Note that unless otherwise stated the equations in this paper are given in SI units.

\section{Atmosphere models\label{drift}}
\begin{figure}
\includegraphics{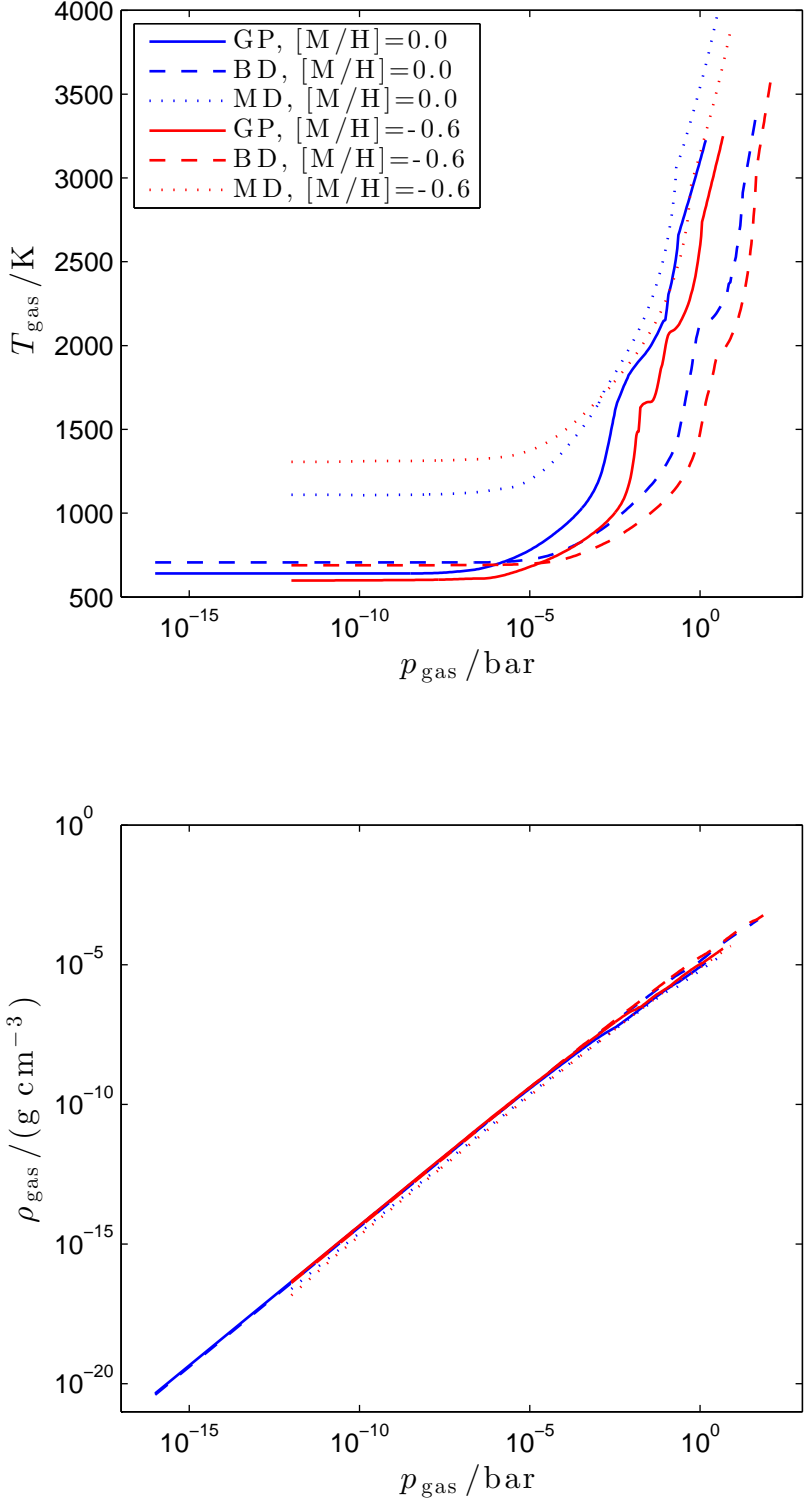}
\caption{Atmospheric $(p_{\rm gas},T_{\rm gas})$ (top plot) and $(p_{\rm gas},\rho_{\rm gas})$ (bottom plot) diagrams for examples of gas giant planets (GP:  $\log{g}=3.0$, $T_{\rm eff}=1500$~K), brown dwarfs (BD:  $\log{g}=5.0$, $T_{\rm eff}=1500$~K) and of M-dwarfs (MD:  $\log{g}=4.0$, $T_{\rm eff}=2700$~K).  These results are obtained from \textsc{Drift-Phoenix} simulations where solar elemental abundances [M/H] $=0.0$ (blue lines) and sub-solar abundances [M/H] $=-0.6$ (red lines) have been considered.  \label{t_p_m}}
\end{figure}

\begin{figure}
\includegraphics{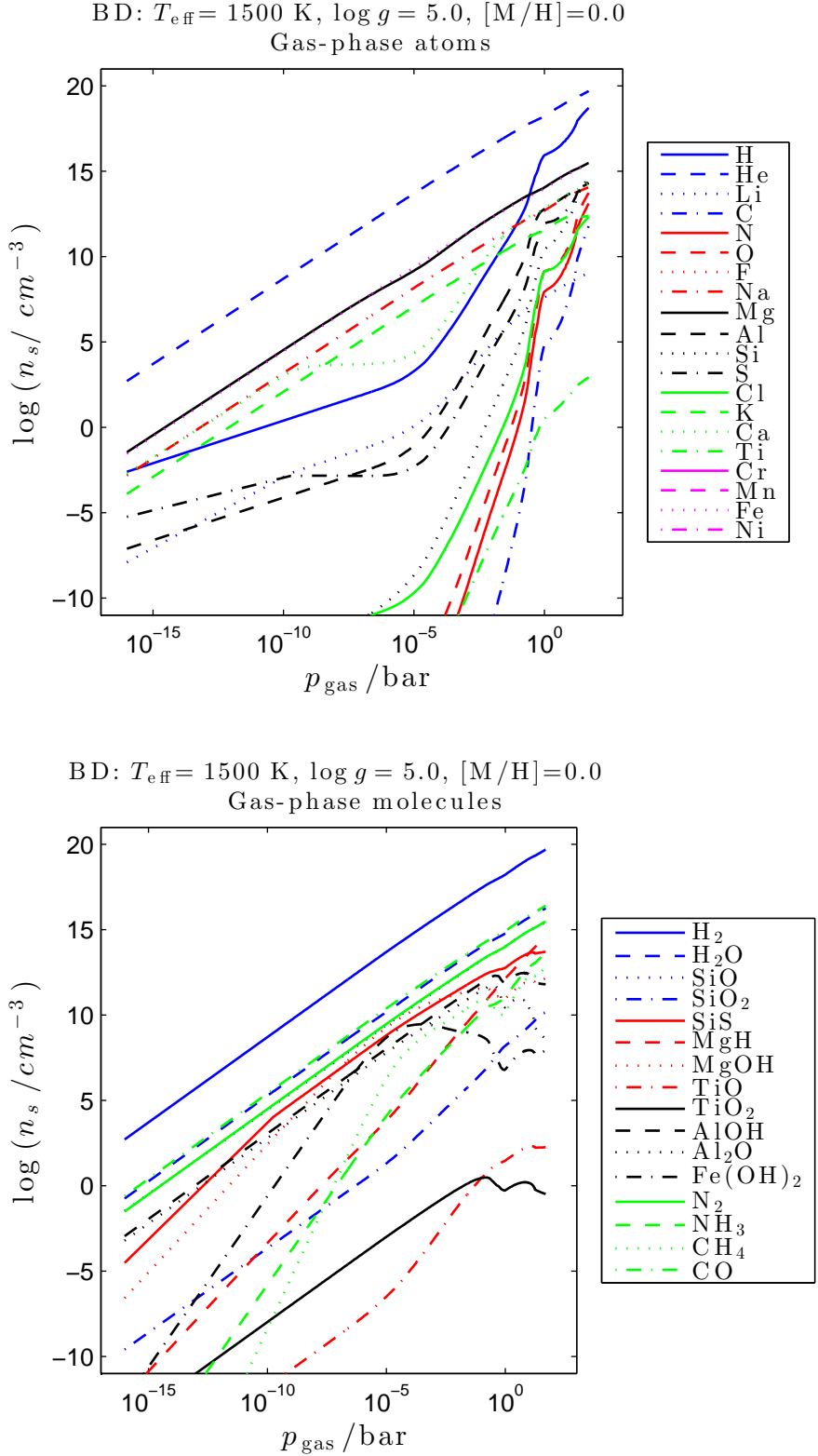}
\caption{Atomic and molecular number densities $n_{s}$ for selected species  $s$ within a brown dwarf (BD: $T_{\rm eff}=1500$~K, $\log{g}=5.0$) model atmosphere assuming initially solar abundances ([M/H] $=0.0$).   For clarity the plots show a close-up of the most abundant species to emphasise their variation.\label{n_bd_s_1}}
\end{figure}

In this paper we are interested in Alfv\'{e}n ionization in the atmospheres of gas giant planets, brown dwarfs and of M-dwarfs.  We utilise the \textsc{Drift-Phoenix} atmosphere grid as described in \cite{witte2011} which provides us with the necessary atmospheric input quantities.  The thermodynamic structure $(p_{\rm gas},T_{\rm gas},\rho_{\rm gas})$ of substellar atmospheres is complex and depends on a number of competing and corroborating factors.  \textsc{Drift-Phoenix} considers an atmosphere in hydrostatic and chemical equilibrium; and utilises mixing length theory and radiative transfer theory to consistently calculate the thermodynamic structure of the model atmospheres.  We consider three representatives sets of stellar parameters for each type of object: giant gas planets ($\log{g}=3.0$, $T_{\rm eff}=1500$~K, hereafter referred to as GP); brown dwarfs ($\log{g}=5.0$, $T_{\rm eff}=1500$~K, hereafter referred to as BD) and M-dwarfs ($\log{g}=4.0$, $T_{\rm eff}=2700$~K, hereafter referred to as MD).  We study solar ([M/H]$=0.0$) and sub-solar ([M/H]$=-0.6$) metallicities.  Fig.~\ref{t_p_m} shows the temperature-pressure profiles and the gas-phase mass density variation with atmospheric pressure for these respective model atmospheres calculated using \textsc{Drift-Phoenix}.  Additionally, as an example, Fig.~\ref{n_bd_s_1} shows the respective atomic and molecular number densities for selected species for a brown dwarf model atmosphere.  

The \textsc{Drift-Phoenix} cloud formation model of Helling and co-workers~\citep{woitke2003,helling2008b,helling2008c} is fundamentally different from the other contemporary cloud models.  First and foremost, their model kinetically describes the formation of cloud particles as a phase-transition process by modelling seed formation; grain growth and evaporation; sedimentation in phase-non-equilibrium; element depletion and the interaction of these collective processes.  Secondly, unlike the other approaches, the trajectory of an ensemble of dust grains is followed from the top of the atmosphere to the bottom.   Additionally, \textsc{Drift-Phoenix} includes the effect of dust formation on the ambient atmosphere in a feedback system that allows the calculation of self-consistent pressure-temperature profiles~\citep{dehn2007}.  In the cloud model the gas phase is calculated assuming chemical equilibrium for $14$ elements (H, He, C, N, O, Si, Mg, Al, S, Na, K, Ti and Ca) and $158$ molecules.  The models initial element abundances are solar (or sub-solar) and are changed by the dust formation process.  The cloud model numerically solves a system of differential equations that describe a 1D stationary dust formation process, where seeds form (TiO$_{2}$) from a highly supersaturated gas, grow to macroscopic particles of micrometer size, gravitationally settle and eventually evaporate as the local temperature exceeds that required for thermal stability.   Once the initial seed particles have formed, growth proceeds via surface chemical reactions that cause the formation of a grain mantle. 
\section{Alfv\'{e}n Ionization in the atmospheres of low-mass objects\label{section2}}
This section discusses Alfv\'{e}n Ionization in the context of substellar atmospheres.  Section~\ref{sec_alf} introduces Alfv\'{e}n ionization and discusses the process by which it can ionize an ambient gas.  The criteria of its applicability is outlined in the context of planetary, brown dwarf and M-dwarf atmospheres (sections~\ref{sec_seed}-\ref{section3}), making use of the \textsc{Drift-Phoenix} model atmospheres discussed in section~\ref{drift}.

\subsection{Alfv\'{e}n Ionization\label{sec_alf}}
\begin{figure}
\includegraphics{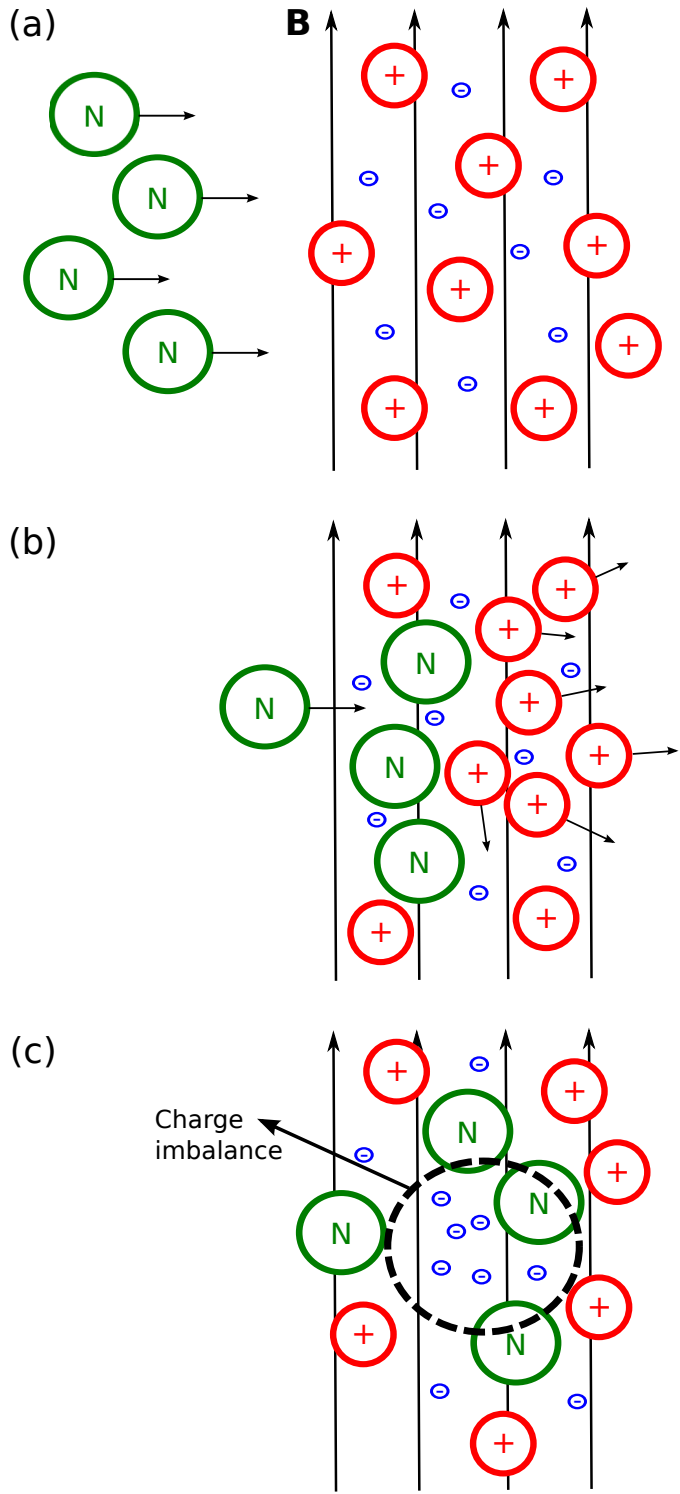}
\caption{Diagram showing the Alfv\'{e}n ionization process:  (a) A constant stream of neutral gas impinges on a low-density localized, magnetized plasma; (b)  The inflowing neutral atoms collide with and displace the plasma ions; (c)  the displaced ions leave behind a significant charge imbalance that accelerates electrons to energies sufficient to ionize the local gas via electron-neutral impact ionization. \label{drawing}}
\end{figure}

Consider a constant stream of neutral gas with a flow speed $v_{0}$ impinging on a low-density stationary, magnetized, seed plasma.  The source of the seed plasma and the required magnetic flux density for the plasma to be sufficiently magnetized will be discussed in section~\ref{sec_seed}~and~\ref{sec_mag}.  It is assumed that this seed plasma is localized by the ambient magnetic field and is of a similar chemical composition as the neutral gas.  The inflowing neutral atoms collide with, and displace the plasma ions from their equilibrium positions producing a significant charge imbalance which cannot be rectified immediately due the restricted motion of the magnetized  electrons (see Fig.~\ref{drawing}).  The scattered ions are sent off in a Larmor orbit, leaving behind them a localized pocket of electrons with a spatial length scale equal to the ion Larmor radius which is defined as $R_{Li}=v_{\perp i}/\omega_{ci}$, where $v_{\perp i}$ is the speed of the ions perpendicular to the magnetic field; and $\omega_{ci}=q_{i}B/m_{i}$ is the ion cyclotron frequency of an ion of mass $m_{i}$ carrying charge $q_{i}$ in a magnetic field.  The resulting self-electrostatic field of the exposed electrons continues to grow until the potential difference inhibits further ionic displacement.  At this point the electrostatic potential energy is now equal to the maximum kinetic energy $\frac{1}{2}m_{\rm gas}v_{0}^{2}$ (where $m_{\rm gas}$ is the mass of a neutral atom) of an ion as a result of a collision~\citep{alfven1960}.  The resulting self-electric field (i.e. the self-repulsion of the electrons) accelerates the local electrons to an equivalent energy, ionizing the incoming neutral atoms that have an electron-impact ionization threshold, $e\phi_{I}$, that is less than  $\frac{1}{2}m_{\rm gas}v_{0}^{2}$~\citep{craig2009,craig2013}.  Note that it is only the relative speed between the magnetized plasma and gas that is important; a moving magnetized plasma can encounter a stationary neutral component: the critical requirement is that the plasma is localized by the ambient magnetic field. The Alfv\'{e}n ionization mechanism is easiest to visualize when there is an interface between the plasma and neutral species; however, the process can occur if the plasma is continuously distributed throughout a volume.  The stochastic nature of collisions is sufficient to scramble the phase and timing of any ion trajectories, leaving the necessary pockets of unbalanced electrons required.  In Alfv\'{e}n ionization laboratory experiments conducted by \cite{fahleson1961}, the plasma and neutral species were mixed and continuously distributed throughout the volume of the vessel.  Many laboratory experiments have successfully exhibited Alfv\'{e}n ionization~\citep{brenning1992,danielsson1973} and many have been performed in space~\citep{newell1985}
For Alfv\'{e}n ionization we require that:  (i)  the seed plasma is strongly magnetized; and (ii) the neutral gas flow reaches a critical speed $v_{c}$.  The critical speed is obtained by equating the kinetic energy of the flow to the electrostatic potential energy of an electron in an electric potential equal to the ionization potential of a neutral gas particle $\phi_{I}$: $\frac{1}{2}m_{\rm gas}v_{c}^{2}=e\phi_{I}$~\citep{alfven1960}.   Therefore, the critical neutral gas flow speed is 
\begin{equation}
v_{c}=\left(\frac{2e\phi_{I}}{m_{\rm gas}}\right)^{1/2}. \label{vcrit}
\end{equation}
Tables~\ref{tbl-1} and~\ref{tbl-2} list the atomic weight, the first ionization potential and the critical ionization speed, calculated using Equation~\ref{vcrit}, for selected atoms and molecules respectively, typically found in the cool atmospheres of low-mass objects such as brown dwarfs, giant gas planets and M-dwarfs.  Atoms such as Potassium (Na) and Chromium (Cr) have the smallest critical speeds ($v_{c}=4.63$ and $5.01$~kms$^{-1}$, respectively) whereas Hydrogen (H) and Helium (He) have the largest speeds ($v_{c}=51.02$ and $34.43$~kms$^{-1}$, respectively).  For a given flow with kinetic energy $E_{k}=\frac{1}{2}m_{gas}v_{0}^{2}$, neutrals with a smaller mass $m_{\rm gas}$ require a greater flow speed $v_{0}$ to  achieve equivalent kinetic energies.  Energetically, condition (ii) is similar to the condition required for the liberation of an electron from the surface of a dust grain (i.e. the work function) as a result of dust-dust or dust-gas collisions~\citep{helling2011b}.

Alfv\'{e}n ionization requires that the local charge imbalance imposed by the gas flow must be established on a timescale, $\tau_{s}$, shorter than that for electron transport to neutralise it, $\tau_{e}$  ~\citep{diver2005}.  Therefore, it is required that $\tau_{s}\ll\tau_{e}$ where: $\tau_{s}$ is the setup time for collisions to create a charge imbalance; and $\tau_{e}$ is the typical time for electron transport to eliminate the charge imbalance.  Consider the case when the ions are magnetized:  $\omega_{ci}\gg\nu_{ni}$, where $\nu_{ni}$ is the ion-neutral collision frequency.  When the ions are scattered by neutral collisions they participate in a Larmor orbit with a frequency $\omega_{ci}$.  As long as the electrons cannot rectify the charge imbalance during this timescale (i.e. before the ions return to their initial positions and neutralize the charge imbalance) then Alfv\'{e}n ionization will occur.  In this case, the critical timescale to create a charge imbalance is the ion cyclotron frequency, $\tau_{s}\approx\omega_{ci}^{-1}$.  However, the ions do not necessarily need to be magnetized for the Alfv\'{e}n ionization process; for example, in~\cite{fahleson1961} the ions are initially unmagnetized.  If unmagnetized, the ions undergo many neutral collisions in an ion cyclotron period, and so are not likely to return as compensating charge to their original equilibrium positions.  This presents a significant relaxation of our more demanding requirement since we do not formally require the ions to be magnetized, only the electrons; hence, the criteria for Alfv\'{e}n ionization is more easily fulfilled.  In this case, the critical timescale would be the rate at which ions are displaced due to neutral collisions: $\tau_{s}\approx\nu_{ni}^{-1}$.  There have been numerous theoretical and numerical studies of the Alfv\'{e}n ionization process that have investigated the ion-neutral beam interaction and the reresulting electron energization~\citep{mcbride1972,machida1986,person1990,mcneil1990}

For gas mixtures, the critical ionization speed $v_{c}$ depends on the relative abundance and critical speeds of the individual participating species.  In experiments conducted using a two-species gas mixtures, $v_{c}$ is intermediate between those of the two individual species.  The behaviour of such multi-species mixtures is not well understood and at best, can be described only by empirical relations~\citep{lai2001}.  For more complex systems, where there is a greater number of participating species (such as in the atmospheres of low-mass objects), we will assume that the individual species will be treated in isolation.
\FloatBarrier
\begin{table}[h!]
\begin{center}
\caption{Table of selected atomic gas-phase species~\citep{diver2005} in substellar atmospheres.  The ionization potential $\phi_{I}$ and the critical ionization speed $v_{c}$ are calculated using Equation~(\ref{vcrit}).  The table is ordered with respect to the magnitude of $v_{c}$, from smallest to largest.  \label{tbl-1}}
\begin{tabular}{lcccc}
\tableline\tableline
Species & Symbol & Atomic weight & $\phi_{I}$ (eV) & $v_{c}$ (kms$^{-1}$) \\
\tableline
Potassium & K& 39.1 & 12.13 & 4.22 \\
Chromium & Cr & 52 & 6.77 & 5.01 \\
Nickel & Ni & 58.69 & 7.64 & 5.01 \\
Manganese & Mn & 54.94 & 7.43 & 5.11 \\
Iron & Fe & 55.85 &7.9 & 5.22 \\
Titanium & Ti & 47.88 & 6.83 & 5.25 \\
Calcium & Ca & 40.08 & 6.11 & 5.42 \\
Aluminium & Al & 26.98 & 5.99 & 6.54 \\
Sodium & Na &22.99 & 5.14 & 6.57 \\
Silicon & Si &28.09 & 8.15 & 7.48 \\
Magnesium & Mg & 24.3 & 7.65 & 7.79 \\
Sulphur & S  & 32.07 & 10.4 & 7.91 \\
Chlorine & Cl & 35.45 & 13 & 8.41 \\
Lithium & Li & 6.94 & 5.39 & 12.24 \\
Oxygen & O & 16 & 13.6 & 12.81 \\
Fluorine & F & 19 & 17.4 & 13.29 \\
Carbon & C & 12.01 & 11.26 & 13.45 \\
Neon & Ne & 20.18 & 21.6 & 14.37 \\
Nitrogen & N & 14.01 & 14.5 & 14.13 \\
Helium & He & 4.0026 & 24.6 & 34.43\\
Hydrogen & H & 1.0079 & 13.6 & 51.02 \\
\tableline
\end{tabular}
\end{center}
\end{table}
\begin{table}[h!]
\begin{center}
\caption{Table of selected molecular gas-phase species in substellar atmospheres.  The ionization potential $\phi_{I}$ and the critical ionization speed $v_{c}$ are calculated using Equation~(\ref{vcrit}). The table is ordered with respect to the magnitude of $v_{c}$, from smallest to largest. \label{tbl-2}}
\begin{tabular}{lccc}
\tableline\tableline
Species & Atomic weight & $\phi_{I}$ (eV) & $v_{c}$ (kms$^{-1}$)  \\
\tableline
TiO & 63.87 & 6.819 & 4.53  \\
TiO$_{2}$ & 79.87 & 9.5 & 4.79   \\
MgOH & 41.31 & 7.5 & 5.91 \\
SiO & 44.08 & 11.3 & 7.03  \\
MgH & 25.31 & 6.9 & 7.25  \\
SiO$_{2}$ & 60.08 & 18 & 7.6 \\
CO & 28.01 & 14.0 & 9.7  \\
N$_{2}$ & 28.01 & 15.6 & 10.4   \\
NH$_{3}$ & 17.04 & 10.02 & 10.65  \\
H$_{2}$O & 18.02 & 12.6 & 11.6 \\
CH$_{4}$ & 16.04 & 12.61 & 12.31 \\
H$_{2}$ & 2.01 & 15.60 & 38.5  \\
\tableline
\end{tabular}
\end{center}
\end{table}
\subsection{Seed plasma\label{sec_seed}}
To satisfy condition (i) for Alfv\'{e}n ionization we require an initial low-density, magnetized seed plasma to initiate the ionization process.  In this section we will compare the length scale over which a plasma can effectively screen a charge imbalance to the length scale over which the Alfv\'{e}n ionization process can create one.  This will determine the required threshold electron number density of the seed plasma needed to trigger Alfv\'{e}n ionization.

As part of the Afv\'{e}n ionization process, the inflowing neutrals scatter the ambient ions exposing a localized pocket of electrons, the self-repulsion of which accelerates the electrons to sufficient energies that collisionally ionize the surrounding neutrals.
Due to the elastic collisions with the neutrals, the scattered ions are given a speed (equal to the neutral flow speed $v_{0}$) perpendicular to the ambient magnetic field and are sent off in a Larmor orbit, leaving behind the magnetized electrons.  The localized pocket of electrons has a spatial length scale equal to the ion Larmor radius, $R_{Li}$.  

For a plasma the Debye length is defined as
\begin{equation}
\lambda_{D}=\left(\frac{\epsilon_{0}k_{B}T_{e}}{n_{e}e^{2}}\right)^{1/2}
\end{equation}
where $T_{e}$ is the electron temperature and $n_{e}$ is the ambient electron number density of the plasma.  The Debye length defines the spatial length scale at which exposed charge is screened by the plasma; at scales greater than the Debye length the plasma is considered neutral and experiences no effects due to the charge imbalance.  At scales less than the Debye length the charge imbalance is not screened and the influence of the resulting electric field is significant.  Therefore, to allow a localized pocket of electrons (a charge imbalance) of length scale $R_{Li}$ to be created it is required that the Debye length must be much smaller than the ion Larmor radius: 
\begin{eqnarray}
\lambda_{D}&\ll&R_{Li}, \\
\left(\frac{\epsilon_{0}k_{B}T_{e}}{n_{e}e^{2}}\right)^{1/2}&\ll&\frac{m_{i}v_{\perp i}}{eB}.
\end{eqnarray}
Rearranging for $n_{e}$, the criterion can be recast in terms of the ambient electron number density of the seed plasma required to initiate Alfv\'{e}n ionization,
\begin{equation}
n_{e}^{\rm seed}\gg\frac{\epsilon_{0}k_{B}T_{e}B^{2}}{m^{2}_{i}v_{\perp i}^{2}}.
\end{equation}
To calculate what threshold electron number density $n_{e}^{\rm seed}$ is required we shall set $v_{\perp i}=v_{0}\approx 1$~kms$^{-1}$ and approximate $m_{i}\approx10^{-27}$~kg.  Although some of the neutral species have a critical speed $O(10~\textnormal{kms}^{-1})$, setting $v_{\perp i}=1$~kms$^{-1}$ gives an upper limit to $n_{e}^{\rm seed}$.  Therefore, the number density of the magnetized seed plasma must satisfy,  
\begin{equation}
n_{e}^{\rm seed}\gtrsim 10^{2}T_{e}B^{2}~~[\textnormal{cm$^{-3}$}].\label{eqn_seed}
\end{equation}
This expression for $n_{e}^{\rm seed}$ is dependent upon the electron temperature $T_{e}$ and the magnetic flux density $B$.  Plasma electron temperatures can range from $T_{e}\approx10^{2}-10^{6}$~K~\citep{fridman2008}, where laboratory microplasmas~\citep{becker2006} and terrestrial lightning strikes can produce $T_{e}\approx10^{4}$~K ($\approx 1$~eV).  In the atmospheres of low-mass objects thermal ionization produces an electron population with $T_{e}\approx10^{2}-10^{3}$~K, and are in thermal equilibrium with the ambient atmospheric gas.  Typical average, global (large-scale) magnetic flux densities for GPs are estimated to be of the order of $\approx 10$~G; with BDs and MDs having much greater flux densities of the order of $1$~kG~\citep{donati2009,reiners2012,christensen2009,sanchez2004,shulyak2011}.  For  $B\approx10-10^{3}$~G and $T_{e}\approx10^{2}-10^{6}$~K, the resulting range in the seed electron number density is $n_{e}^{\rm seed}\gtrsim10^{6}-10^{16}$~cm$^{-3}$.  For a typical GP atmosphere with $T_{e}\approx O(1000$~K) and $B\approx O(10$~G) the threshold seed electron number density is $n_{e}^{\rm seed}\gtrsim10^{7}$~cm$^{-3}$.  For BD and MD atmospheres with a similar order of magnitude electron temperature ($T_{e}\approx O(1000$~K)) and $B\approx O(1$~kG), the threshold seed density is $n_{e}^{\rm seed}\gtrsim10^{11}$~cm$^{-3}$.

\begin{figure}
\includegraphics{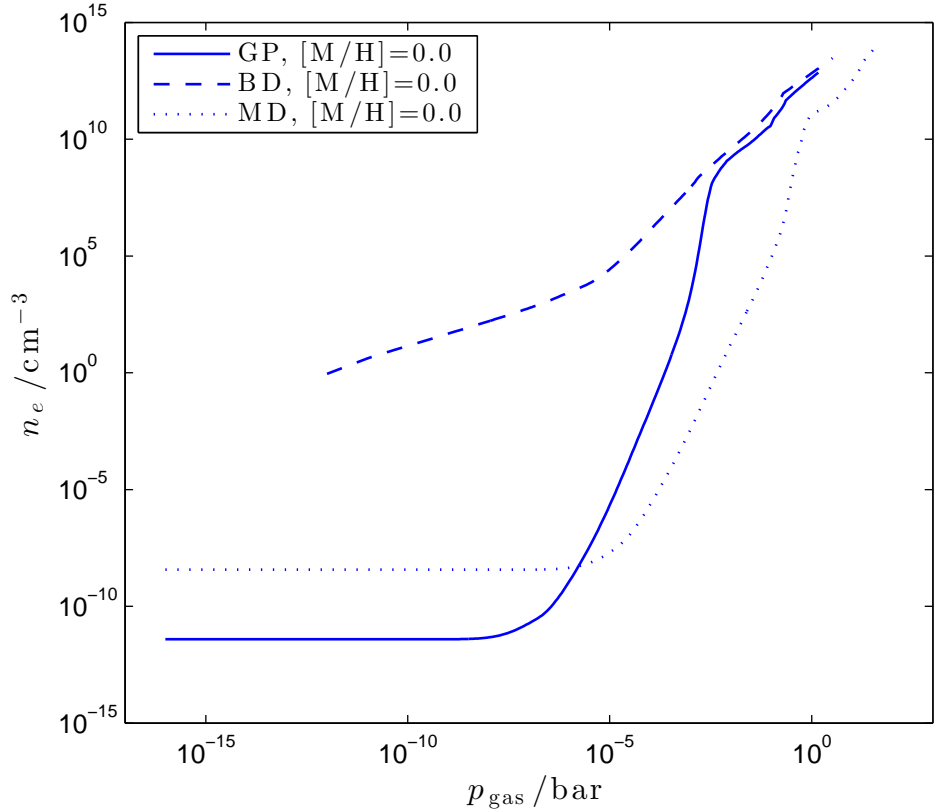}
\caption{The ambient electron number density $n_{e}$ as a function of atmospheric pressure $p_{\rm gas}$.  The electron number density results from thermal ionization in the atmosphere of a gas giant planet (GP:  $\log{g}=3.0$, $T_{\rm eff}=1500$~K), brown dwarf (BD:  $\log{g}=5.0$, $T_{\rm eff}=1500$~K) and M-dwarf (MD:  $\log{g}=4.0$, $T_{\rm eff}=2700$~K) for solar abundance ([M/H] $=0.0$). \label{ne_p}}
\end{figure}

In the atmospheres of low-mass objects, ambient electron number densities from thermal ionization are expected to reach up to $n_{e}\approx10^{12}$~cm$^{-3}$ deep in the atmosphere ($p_{\rm gas}\approx1$~bar) (see Fig.~\ref{ne_p}).  However, at higher altitudes, the atmospheric pressure and temperature decrease and the gas becomes insufficiently energetic to maintain the same level of thermal ionization and hence $n_{e}$ decreases.  As an example, for a BD atmosphere with [M/H]=0.0, $n_{e}$ falls from $\approx10^{12}$~cm$^{-3}$ (at $p_{\rm gas}\approx1$~bar) to $\approx10$~cm$^{-3}$ (at $p_{\rm gas}\approx10^{-12}$~bar).  
\begin{figure}
\includegraphics{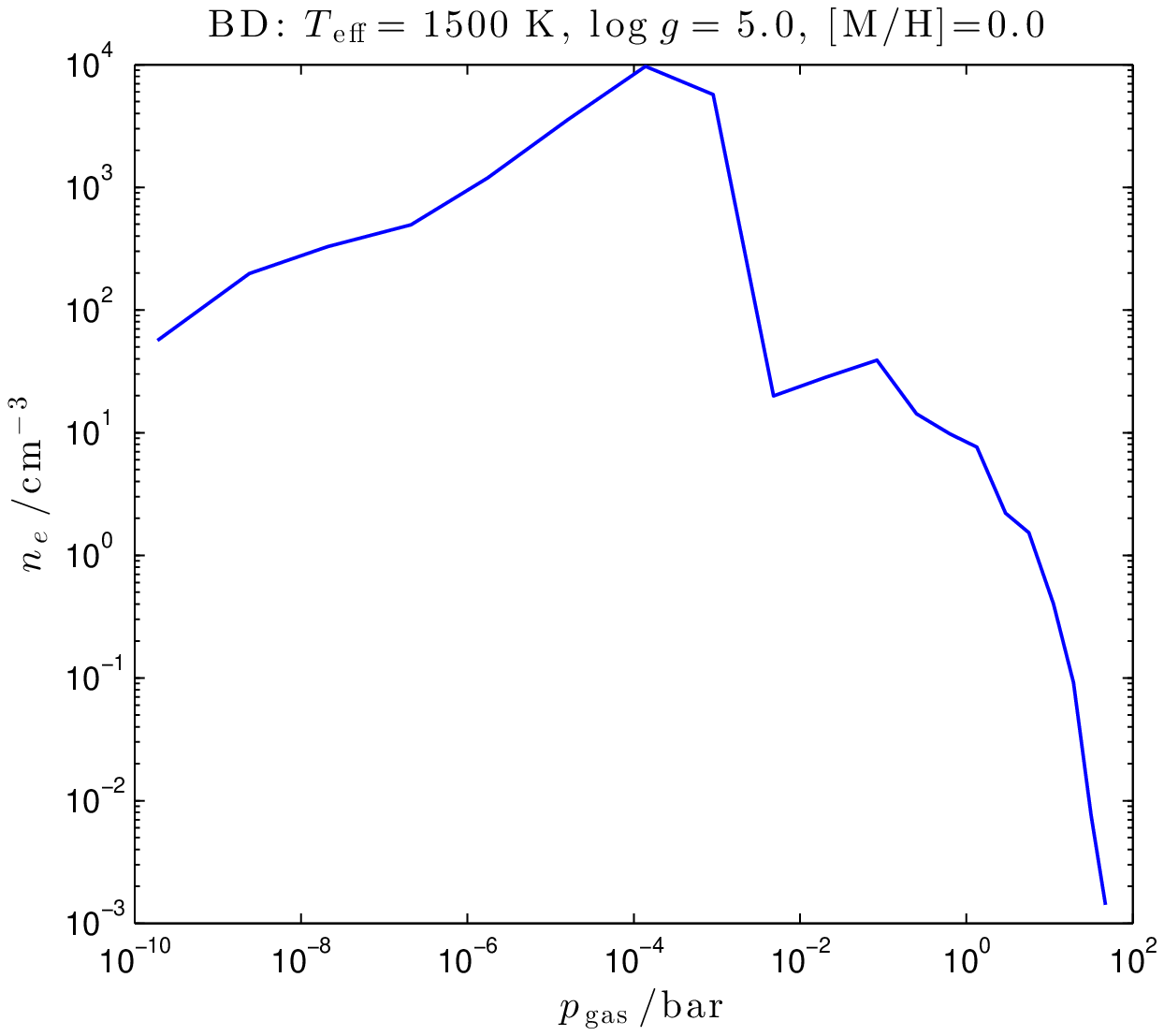}
\caption{Electron number density $n_{e}$ resulting from the cosmic-ray bombardment of a brown dwarf atmosphere (BD:  $\log{g}=5.0$, $T_{\rm eff}=1500$~K, [M/H]=0.0).  \label{cos_ray}}
\end{figure}
Fig.~\ref{ne_p} shows the electron number density as a function of atmospheric pressure due to thermal ionization alone (resulting from Saha equilibrium).  Although $n_{e}$ falls to low values in the upper regions of the atmosphere, we expect that the electron number density will be enhanced via other ionizing processes such as cosmic-ray ionization and atmospheric inter-grain electrical discharges (there are other processes too) locally yielding an electron number density that exceeds the threshold value required for the seed plasma.  Clouds in low-mass objects are susceptible to local discharge events (lightning) which can significantly enhance the local electron density~\citep{helling2011a,helling2013} e.g. observations of terrestrial lightning strikes show the electron number density $n_{e}\approx10^{17}$~cm$^{-3}$ in a single discharge event~\citep{uman1964,guo2009,chang2010}.  Such lightning strikes are analogous to gas discharges used in terrestrial laboratory plasmas, such as microdischarge devices, where characteristic plasma number densities can attain $\approx10^{16}$~cm$^{-3}$~\citep{becker2006}. 

Additionally, cosmic ray bombardment of the atmosphere will significantly increase $n_{e}$ via impact ionization of the initially neutral atmosphere. Fig.~\ref{cos_ray} shows the resulting electron number density $n_{e}$ from the bombardment of a brown dwarf atmosphere (BD, [M/H] $=0.0$) by cosmic rays.  The plot shows that high in the atmosphere ($p_{\rm gas}\approx10^{-4}$~bar) cosmic ray ionization processes can increase the ambient electron number density to $n_{e}\approx10^{4}$~cm$^{-3}$.  For more details regarding the cosmic-ray calculation see Appendix~\ref{c_ray}.

Taking into account these additional contributions to the electron number density, it is reasonable to presume that the seed plasma required for Alfv\'{e}n ionization is achievable in the atmospheres of low-mass objects such as M-dwarfs, brown dwarfs and gas planets. 

\subsection{Magnetized seed plasma\label{sec_mag}}
Furthermore, to satisfy condition (i) for Alfv\'{e}n ionization the background seed plasma must be magnetized.  Since Alfv\'{e}n ionization works when the electrons are magnetized, whether or not the ions are too, ensuring that the electrons are magnetized is sufficient.  The criterion for magnetized electrons is $\omega_{ce}\gg\nu_{coll}$, where $\omega_{ce}=eB/m_{e}$ is the electron cyclotron frequency; $m_{e}$ is the mass of an electron; and $\nu_{coll}$ is the electron-neutral collision frequency.  This simply states that for the plasma dynamics to be significantly influenced by a magnetic field, a charged particle must complete on average at least one gyration before participating in a collision.  The criterion for magnetised electrons can be rewritten as 
\begin{eqnarray}
\frac{eB}{m_{e}}&\gg&n_{\rm gas}\langle\sigma v \rangle, \\
B&\gg&\frac{m_{e} n_{\rm gas}\langle\sigma v \rangle}{e}, \\
&\approx&\frac{m_{e} n_{\rm gas}\pi r_{0}^{2}}{e}\left(\frac{k_{B}T_{e}}{m_{e}}\right)^{1/2}, \label{eqn_mag}
\end{eqnarray}
where the approximations $\langle \sigma v \rangle\approx\pi r^{2}_{0}\langle v\rangle$ and $\langle v\rangle=(k_{B}T_{e}/m_{e})^{1/2}$ have been made; and $r_{0}$ is the atomic radius.  \textsc{Drift-Phoenix} simulations provide the atmospheric pressure and temperature structure ($p_{\rm gas}$, $T_{\rm gas}$) for GP, BD and MD atmospheres (Figs.~\ref{t_p_m}~and~\ref{n_bd_s_1}).  The required local magnetic flux density $B$ needed to establish a magnetized plasma for Alfv\'{e}n ionization can be calculated for each of these atmospheres.  

\begin{figure}
\includegraphics{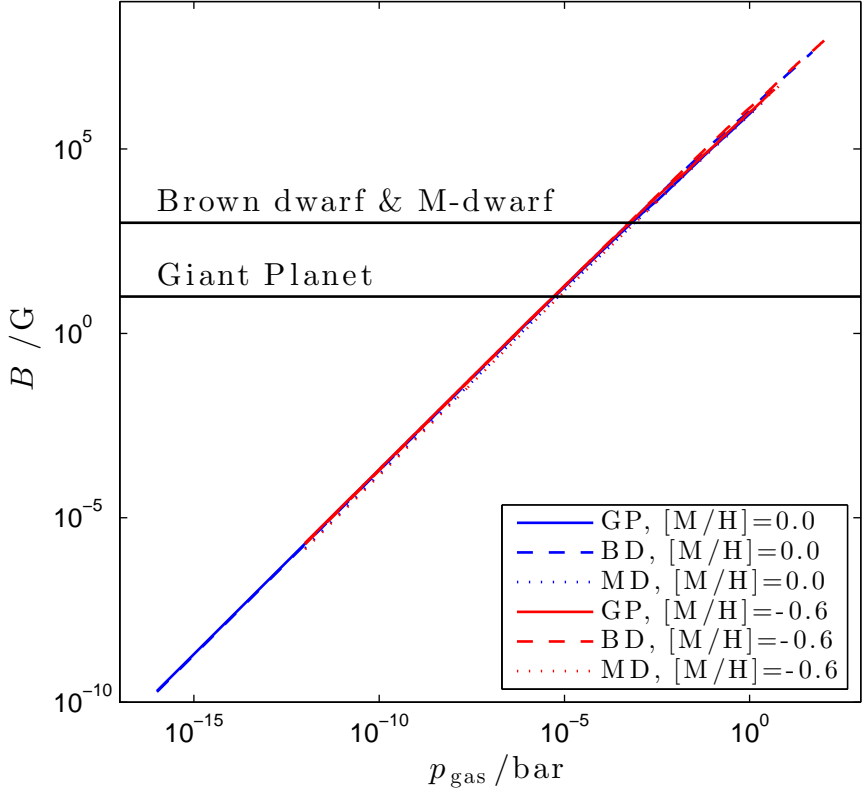}
\caption{The magnetized plasma criterion (Eqn.~\ref{eqn_mag}) in the atmospheres of low-mass objects: gas giant planets (GP:  $\log{g}=3.0$, $T_{\rm eff}=1500$~K; solid line); brown dwarfs (BD:  $\log{g}=5.0$, $T_{\rm eff}=1500$~K; dashed line) and M-dwarfs (MD:  $\log{g}=4.0$, $T_{\rm eff}=2700$~K; dotted line) for both solar abundance ([M/H] $=0.0$, blue plots) and low metallicity ([M/H] $=-0.6$, red plots). The plot shows the required minimum magnetic flux density $B$ for the plasma to qualify as magnetized.  The horizontal black lines show the typical average, global (large-scale) magnetic flux densities for GP, BD and MD. \label{b_p}}
\end{figure}

Fig.~\ref{b_p} shows the criteria for a magnetized plasma (Eqn~\ref{eqn_mag}) in the atmospheres of low-mass objects: GP, BD and MD atmospheres for both [M/H] $=0.0$ and [M/H] $=-0.6$.  In these calculations the atomic radius has been approximated as $r_{0}\approx10^{-8}$~cm and $T_{e}\approx T_{\rm gas}$.  For a typical atmosphere such as for a brown dwarf (BD, [M/H] $=0.0$: see Figs.~\ref{t_p_m} and~\ref{n_bd_s_1}) with $p_{\rm gas}\approx10^{-10}$~bar, $n_{\rm gas}\approx10^{8}$~cm$^{-3}$ and $T_{e}\approx 700$~K then the magnetized plasma criterion is $B\gtrsim10^{-4}$~G.  This means that for the plasma to be considered magnetized the ambient magnetic flux density must be at least $10^{-4}$~G.  

Fig.~\ref{b_p} shows that very little discriminates between the criteria for the various atmospheric models.  Only deep in the respective atmospheres, where the gas number density is high, is a significant magnetic flux density required to satisfy the criterion.  As the gas density increases, the plasma electrons are more likely to collide with the neutrals inhibiting their gyration and so disrupt the influence of the ambient magnetic field on their motion.  High in the atmosphere ($p_{\rm gas}\approx 10^{-12}$~bar) $B\gtrsim10^{-6}$~G is required, whereas deep in the atmosphere the criterion is much more severe $B\gtrsim10^{6}$~G.  However, over a large atmosphere pressure range ($10^{-15}\lesssim p_{\rm gas}\lesssim10^{-2}$~bar) the magnetized plasma criterion is easily achievable and the minimum local magnetic flux density required is $\lesssim10^{4}$~G.  Note that GP and BD atmospheric models for [M/H]$=0.0$ extend to lower atmospheric pressures ($p_{\rm gas}\approx 10^{-16}$~bar) where the magnetized criterion is $B\gtrsim10^{-10}$~G; additionally BD models for both [M/H]$=0.0$ and $-0.6$ can achieve $p_{\rm gas}\approx10$~bar deep in the atmosphere where $B\gtrsim10^{8}$~G.

Typical average, global (large-scale) magnetic flux densities for GPs are estimated to be of the order of $\approx 10$~G; with BDs and MDs having much greater flux densities of the order of $1$~kG~\citep{donati2009,reiners2012,christensen2009,sanchez2004,shulyak2011}.  However, note that stronger, localized (small-scale) magnetic flux densities maybe likely; for example, the Sun's dipole magnetic flux density (large-scale field) has a magnitude $\approx O(10~\textnormal{G})$, but small-scale fields, e.g. a sunspot, can have magnitudes $\approx O(1~\textnormal{kG})$~\citep{priest1985,borrero2008}.  Although the low-mass objects we consider are physically different from the Sun;  the principle still applies that local fluctuations in the mean field can lead to favourable enhancements in field strength of the order required for optimum performance in our description.  Therefore it can be expected that the seed plasma is sufficiently magnetized in the majority of the atmosphere and can be considered a magnetized plasma.  However, such local enhancements of the magnetic flux density may not be readily achievable for GPs since the source of magnetic amplification, expected in BDs and MDs, may be absent.  In addition, GPs magnetic flux density is expected to be much weaker in strength ($10$~G) in comparison to BDs and MDs ($1$~kG), as a result Alfv\'{e}n ionization is harder to achieve in GPs.  However, the nature of exoplanetary magnetic fields is still open for discussion and further investigation, e.g. the enhancement of the magnetic field strength in exoplanets can be achieved through the formation of magnetodisks~\citep{khoda2012}.

We have assumed that the seed plasma has had sufficient time to come into thermal equilibrium with the local gas and can be approximated to have a similar temperature, $T_{e}\approx T_{\rm gas}$.  In the {\sc Drift-Phoenix} simulations the electrons liberated by thermal ionization are in thermal equilibrium with the ambient gas and so this approximation is valid; however, this may not be the case for plasma electrons produced from other processes such as gas discharges.  In reality the plasma electrons need not be in thermal equilibrium with the ions or neutrals (a non-thermal plasma) and their temperature may be lower or much higher than these species~\citep{fridman2008}.  For example, laboratory plasma discharges can have electron temperatures $T_{e}\approx1-10$~eV (equivalent to $10^{4}-10^{5}$~K, \cite{diver2001,fridman2008}) and can vary between $0.01-1$~eV (equivalent to $10^{2}-10^{4}$~K) in the Earth's ionosphere~\citep{fridman2008}.  The effect of having a greater electron temperature would be to increase the minimum threshold magnetic flux density for the atmospheric plasma to be considered magnetize, for example if $T_{e}\approx1$~eV (or a higher temperature, $T_{e}\approx100$~eV) the minimum magnetic flux density required would increase by a factor $\approx 3$ ($\approx30$) relative to the case $T_{e}=1000$~K ($T_{e}\approx0.1$~eV).  This effect is model independent since we are discussing an additional population of electrons that arise from other ionising processes than those calculated using the {\sc Drift-Phoenix} atmospheric model.
\subsection{Minimum electron number density for Alfv\'{e}n Ionization}
 In Alfv\'{e}n ionization a local pocket of unscreened electrons (a charge imbalance) is produced by the inflowing neutrals. The self-electric field of the electrons (their collective electrostatic self-repulsion) accelerates the electrons to energies that can ionize the surrounding neutrals via electron-impact ionization.  In order for the electrons to ionize the neutrals, the number density of the unscreened electron pocket must be great enough to facilitate the acceleration of the electrons to the appropriate ionization energy.  The pocket of electrons will have a net charge $Q$, and a corresponding electrostatic potential, with a length scale equal to the ion Larmor radius $R_{Li}$.  For these electrons to successfully ionize a neutral gas with an ionization potential $\phi_{I}$, the electrostatic potential of the localized pocket must equal $\phi_{I}$:
\begin{equation}
\phi_{I}=\frac{Q}{4\pi\epsilon_{0}R_{Li}}
\end{equation}
Furthermore, the volume of the charge pocket is $\approx R_{Li}^{3}$ hence $Q\approx en_{e}R_{Li}^{3}$; therefore,
\begin{equation}
\phi_{I}=\frac{n_{e}eR_{Li}^{2}}{4\pi\epsilon_{0}}.
\end{equation}
Therefore, the minimum electron number density $n_{e}^{\rm min}$ of the charge imbalance required to produce the self-field for electron-impact ionization can be calculated,
\begin{equation}
n_{e}^{\rm min}\geq\frac{4\pi\epsilon_{0}\phi_{I}Z_{i}^{2}eB^{2}}{m_{i}^{2}v_{\perp i}^{2}}, \label{nemin}
\end{equation}
where $R_{Li}=m_{i}v_{\perp i}/(q_{i}B)$ is the Larmor radius, $q_{i}=Z_{i}e$ and $Z_{i}$ is the ion charge number.  We will assume that $Z_{i}=1$, $v_{\perp i}=v_{c}=(2e\phi_{I}/m_{\rm gas})^{1/2}$, and $m_{i}\approx m_{\rm gas}$, then
\begin{equation}
n_{e}^{\rm min}\geq\frac{2\pi\epsilon_{0}B^{2}}{m_{\rm gas}}.
\end{equation}   
For a GP with a magnetic flux density of $10$~G and a H$_{2}$ dominated gas, the minimum electron number density to produce the required self-field is $n_{e}^{\rm min}\gtrsim10^{4}$~cm$^{-3}$.  In the atmosphere of a BD or MD where magnetic flux densities can be $\approx1$~kG, for a H$_{2}$ dominated gas $n_{e}^{\rm min}\approx10^{8}$~cm$^{-3}$.

\begin{figure}
\includegraphics{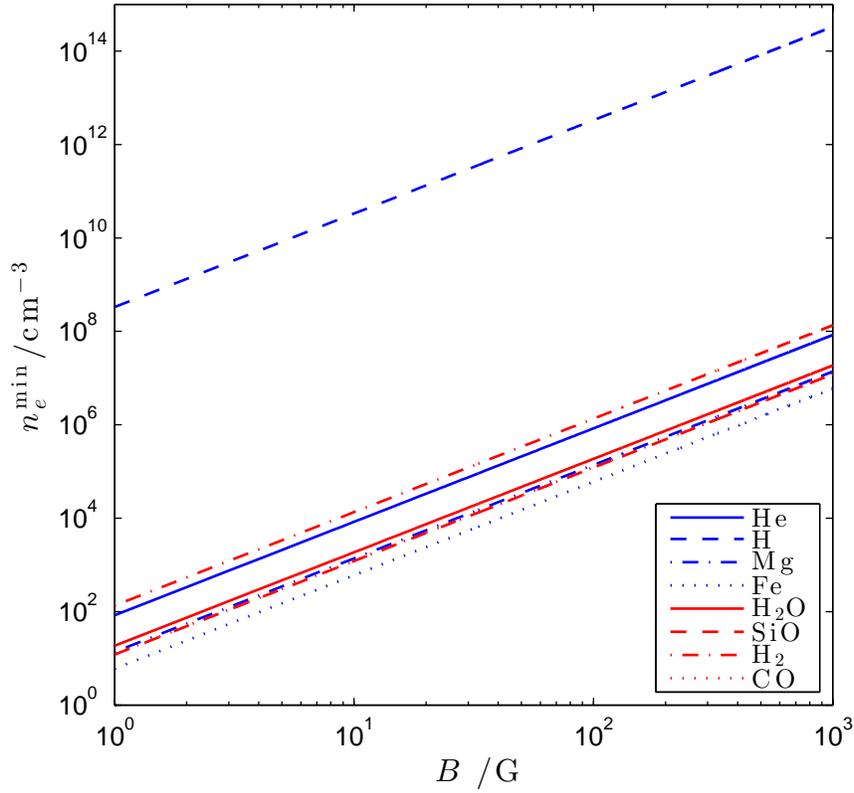}
\caption{The minimum electron number density $n_{e}^{\rm min}$ (Eqn.~\ref{nemin}) required to produce the self-field for electron-impact ionization to occur for selected gas phase species.  The top plot shows $n_{e}^{\rm min}$ as a function of the magnetic flux density $B$ for $T_{i}=500$~K for selected gas phase species; the bottom plot shows how $n_{e}^{\rm min}$ varies with the atmospheric gas $(p_{\rm gas},T_{\rm gas})$ for Hydrogen and a constant magnetic flux density $B=100$~G.
\label{ne_b}}
\end{figure}

Fig.~\ref{ne_b} shows the minimum electron number density $n_{e,}^{\rm min}$ as a function of magnetic flux density $B$ for a range of gas phase species.  In this approximation the species dependence on $n_{e}^{\rm min}$ is through the term $m_{i}^{-1}$, where species with a small mass require a larger $n_{e}^{\rm min}$, hence Hydrogen requires a greater $n_{e}^{\rm min}$ than Iron.  For example, the minimum electron number density for a Hydrogen (Iron) dominated plasma when $B\approx10^{2}$~G is $n_{e}^{\rm min}\geq10^{12}$~cm$^{-3}$ ($\geq10^{5}$~cm$^{-3}$).  As $B$ is increased, $R_{Li}$ decreases as the ions are forced to participate in smaller orbits.  As a result, so that the same electric potential ($\phi_{I}$) can be maintained, $n_{e}^{\rm min}$ must increase since $\phi_{I}\propto n_{e}^{\rm min}R_{Li}^{2}$.
\subsection{Metastable states of gas-phase species\label{sec_meta}}
In general, if the electron-neutral impact energy is insufficient to attain the ionization energy of the target particle, the energy maybe sufficient that the bound electrons are excited into higher electronic states, and subsequently return to a lower energy state, by radiating a photon.  The timescale of such an electronic transition is usually of the order of $10^{-8}$~s~\citep{bransden2000}; however, certain neutral atoms and molecules have excited states that persist for longer timescales (see Table~\ref{meta}).  Such atoms and molecules struggle to decay to lower energy states because they cannot relax via dipole radiation and so must do so via higher-order transitions which are less efficient.  Such {\it metastable states} can have lifetimes of the order of seconds~\citep{dunning1996} and can have a big impact on the distribution of energy within substellar atmospheres.  Metastable transitions have been observed in the ejecta of $\eta$ Carinae~\citep{zethson2001,hartman2005}

Metastable states can play a crucial role in Alfv\'{e}n ionization, where metastable species can interact with another energetic electron, this time imparting enough energy to ionize the excited species.  The advantage being that the ionising electron does not need to transfer an energy equivalent to the ionization energy of the species, only the difference in energy between the metastable state and the ionization energy, thus requiring a lower Alfv\'{e}n ionization critical speed.  This means that for metastable species such as Helium  or H$_{2}$, the species can be ionized in two stages without the total critical velocity ($v_{c}=34.43$~kms$^{-1}$ and $v_{c}=38.5$~kms$^{-1}$, respectively) needing to be reached, with a lower speed being sufficient.  Hence metastables are easier to ionize.  Table~\ref{meta} gives a selection of metastable states appropriate to the atmospheres considered here.  

The required critical speed for some metastable states is still quite tough to achieve; for example, metastable state 2$^{3}$S$_{1}$ of He requires $v_{c}=30.9$~kms$^{-1}$ (see Table~\ref{meta}).  However, the critical speed $v_{c}$ is a bulk fluid commodity, averaged over a particle energy distribution function; in reality, only a high energy tail of the distribution function is required to attain these metastable states and the bulk, average speed for the remaining energy to fully ionize the neutral.  Hence, it may be enough to have the mean fluid speed much smaller than $v_{c}$ since the energetic tail of the underlying particle distribution function might be populated by enough particles to excite the metastable state initially.

Furthermore, ionization can also occur through the collision of metastable neutrals ($X^{*}$) with neutral particles of another species ($Y$) if the excited metastable state has an energy equal to or greater than the ionization potential of the neutral atom.  This is known as the Penning Ionization Effect~\citep{fridman2008}:
\begin{equation}
X^{*}+Y\rightarrow X+Y^{+}+e^{-}.
\end{equation}      
As an example, consider metastable He ($19.82$~eV) and neutral Na ($\phi_{I}=12.13$~eV), should they collide the latter will be ionize and the former will return to its ground state.  An observable consequence of the Penning Effect will be the presence of forbidden lines in observed spectra, not present in the non-ionising case.  

\begin{figure}
\includegraphics{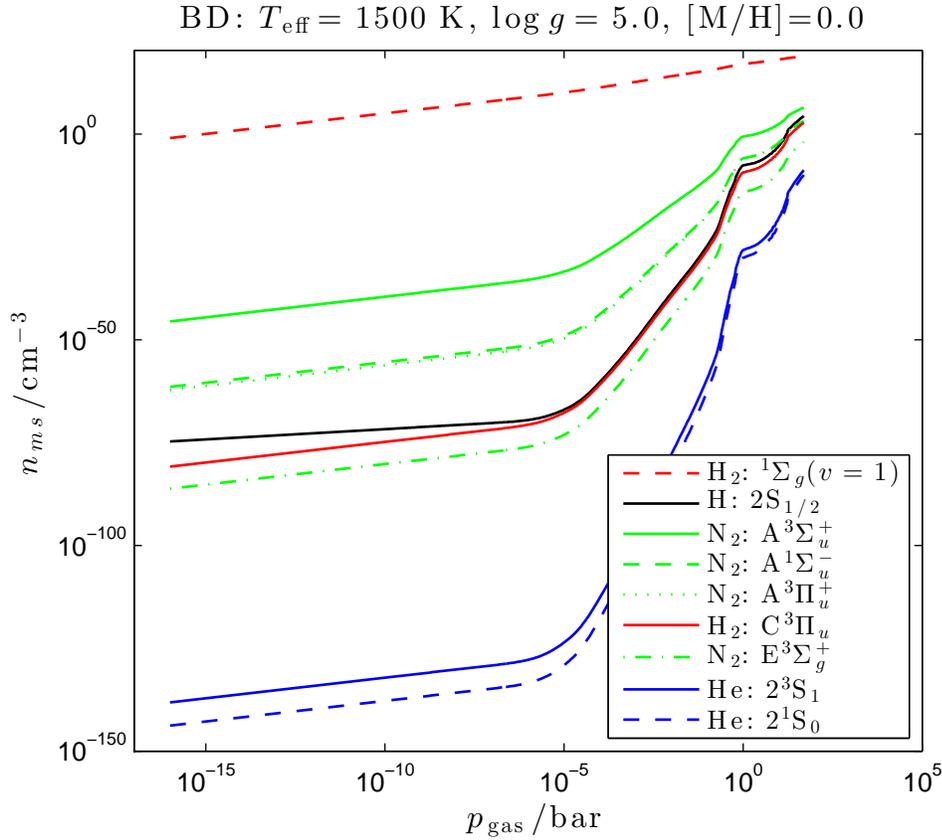}
\caption{The number density of selected metastable states $n_{ms}$ for a brown dwarf (BD:  $\log{g}=5.0$, $T_{\rm eff}=1500$~K, [M/H]=0.0) calculated using Eqn.~\ref{meta_eqn}.  Information regarding the metastable states can be found in Table~\ref{meta}. \label{meta_fig}}
\end{figure}

It is useful to quantify what proportion of the atmosphere is in a metastable state.  Assuming the system is in thermodynamic equilibrium with temperature $T$ and that it can be described by a canonical ensemble, the probability that the system is in state $E_{n}$ is given by
\begin{equation}
p_{n}=\frac{g_{n}\exp{(-E_{n}/(k_{B}T))}}{\sum_{i}g_{i}\exp{(-E_{i}/(k_{B}T))}}
\end{equation}
where $g_{j}$ is the degerancy factor of the microstate $j$ and the sum is over all energy levels.  Therefore, the probability ratio of the system being in the ground state (gs) relative to the metastable state (ms) $p_{ms}/p_{gs}$, can be calculated.  Hence the relative number density $n_{ms}$ of Helium that exists in the metastable state with energy $E_{ms}$ and degeneracy factor $g_{ms}$ is given by
\begin{equation}
n_{ms}=n_{gs}\frac{g_{ms}}{g_{gs}}\exp{\left(-\frac{\Delta E}{k_{B}T}\right)} \label{meta_eqn}
\end{equation}
where $n_{gs}$ is the number density of the species in the ground state; $g_{gs}$ is the degeneracy factor of the ground state; and $\Delta E$ is the energy difference between the ground state $E_{gs}$ and the metastable state $E_{ms}$.  Fig.~\ref{meta_fig} shows the number density of metastables in a brown dwarf atmosphere (BD, [M/H]=0.0) as a function of atmospheric pressure and temperature $(p_{\rm gas}, T_{\rm gas})$, for selected metastable species (listed in Table~\ref{meta}).  We consider the metastable states of H, He and H$_{2}$ primarily because they are the hardest to ionize (i.e. they have the largest critical speed) via Alfv\'{e}n ionization.  He and H$_{2}$ are of further interest since they have the largest number density in the atmosphere; and N$_{2}$ is interesting from a thermal equilibrium perspective since it is a very stable molecule.  Thermally, there is insufficient energy in the atmosphere to excite the selected species into their metastable states to produce a significant number density $n_{ms}$:  metastable He is totally negligible.  The metastable states of molecular nitrogen (N$_{2}$) yield higher number densities ($n_{ms}\approx10^{-15}$~cm$^{-3}$) but not significantly so.  One metastable state of H$_{2}$ ($^{1}\Sigma_{g}(v=1)$) yields a modest $n_{ms}\approx10^{10}$~cm$^{-3}$ at $p_{\rm gas}\approx1$~bar; however, the energy of the microstate is very small (0.52~eV) that its contribution to the overall Alfv\'{e}n ionization process would be insignificant.

In the atmospheric models considered here, the ambient atmosphere does not harbour a significant population of metastable species produced from thermal excitation.  For an atmosphere to have a significant population of metastables the atmospheric temperature and/or the density of the ground state species (corresponding to the metastable species) would have to be greater.  In the model atmosphere results considered here, excitation to metastable states would need to occur via an alternative process such as gas discharges; cosmic-ray interactions; or even as part of the Alfv\'{e}n ionization mechanism where bombardment from neutral flow drives the excitation.  For example, simulations of electron avalanches (the precursor to a full gas discharge) in 1 mm gap of atmospheric Nitrogen have shown that the metastable density can reach $\approx10^{12}$~cm$^{-3}$~\citep{craig2013}.  Furthermore, if in a localized volume of a BD, [M/H]$=0.0$ model atmosphere  (for example at an atmospheric pressure of $\approx10^{-10}$~bar), Alfv\'{e}n ionization excited all the Helium from its ground state to its first metastable state, the metastable density would be $\approx10^{10}$~cm$^{-3}$ (see Fig.~\ref{n_bd_s_1}).
\begin{table}
\begin{center}
\caption{Selected metastable states that may play a significant role in Alfv\'{e}n ionization and the chemistry occurring in substellar atmospheres.~\citep{dunning1996}\label{meta}}
\begin{tabular}{lcccc}
\tableline\tableline
Species & State & Energy (eV) & Approximate lifetime (s) & $v_{c}$ (kms$^{-1}$)\\
\tableline
H  & 2S$_{1/2}$ & 10.19 & 0.15 & 44.14\\
\tableline
He & 2$^{3}$S$_{1}$ & 19.82 & 7.9$\times$10$^{3}$ & 30.9 \\
 & 2$^{1}$S$_{0}$ & 20.61 & 2$\times$10$^{-2}$ & 31.5 \\
\tableline
H$_{2}$ & C$^{3}\Pi_{u}$  & 11.75 & 1$\times$10$^{-3}$ & 33.56\\
  & $^{1}\Sigma_{g}(v=1)$ & 0.52 & 4$\times$10$^{6}$ & 7.06\\
\tableline
N$_{2}$ & A$^{3}\Sigma^{+}_{u}$ & 6.17 &  2.7 & 6.52\\
&  A$^{1}\Sigma^{-}_{u}$ &  8.40 &  0.7 & 7.6\\
& A$^{3}\Pi^{+}_{u}$ &  8.55 &  1.7$\times$10$^{-4}$ & 7.67\\
& E$^{3}\Sigma^{+}_{g}$ & 11.87 &  2$\times$10$^{-4}$ & 9.04\\
\tableline
\end{tabular}
\end{center}
\end{table}

\subsection{Atmospheric flows and the critical speed $v_{c}$\label{section3}}

For Alfv\'{e}n ionization condition (ii) must be satisfied i.e. the neutral gas flow speed must reach the critical flow speed $v_{c}$ which, for most gas species of interest $v_{c}\approx O(1-10$~kms$^{-1})$.  It is therefore natural, in the context of neutral gas atmospheres around planetary and substellar objects, to look at dynamic meteorology and atmospheric flows to ascertain if such flow speeds are plausible.  

In the absence of significant ionization, the atmospheric seed plasma is a minority species relative to the neutral species if thermal ionization is considered only.  Therefore, the dynamics of the atmospheric gas-plasma mixture will be dominated by that of the neutral gas.  Note this does not mean that the plasma component is not magnetized (or coupled to the magnetic field) but rather that the dynamical effect of the plasma is a second order effect relative to the neutral species.  We are proposing Alfv\'{e}n ionization as a process to significantly enhance the degree of ionization and boost the plasma population, allowing the subsequent use of continuum models (such as magnetohydrodynamics, cold or warm plasma theory, etc).  With this in mind it is reasonable to omit plasma behaviour for now and focus on neutral gas flows as a starting point (which are unaffected by the magnetic field) to see if we can create a significant plasma via Alfv\'{e}n ionization.  However, if an atmospheric plasma can be established (via Alfv\'{e}n ionization) and sustained the effect of the magnetic field on flows will be important and should be incorporated into the governing equations of motion.

There has been significant progress in the meteorological studies of extrasolar planetary atmospheres, and by analogy brown dwarfs, with sophisticated numerical simulations of global atmospheric circulatory systems~\citep{showman2002,showman2008,showman2009,menou2009,rauscher2010,cooper2005,dobbs2008,dobbs2010,dixon2012,lewis2010,heng2011}.  

The atmospheric dynamics of Hot Jupiters are of particular interest because of their high temperatures and short orbital periods.  Three-dimensional simulations of the atmospheric circulation on hot Jupiters HD 209458b and HD 189744b  by~\cite{showman2002,showman2008,showman2009} globally solve the primitive equations governing the atmospheric gas dynamics and compare the infrared spectra and light curves predicted by the simulations to observational data.  Their models produce equatorial jets with flow speeds of $3-4$~kms$^{-1}$, which are of the order of magnitude required to trigger Alfv\'{e}n ionization.~~\cite{menou2009} solve the primitive equations  using a pseudospectral solver for a shallow three-dimensional hot Jupiter model of HD 209458b and find that in the barotropic regime, zonal jets develop that obtain speeds up to $4$~kms$^{-1}$ and up to $10$~kms$^{-1}$~\citep{rauscher2010}, this is consistent with the meteorological simulations of~\cite{cooper2005}.  Dobbs-Dixon and co-workers have incorporated radiative transfer and viscous effects into their hydrodynamical simulations yielding gas flows $\approx 5$~kms$^{-1}$~\citep{dobbs2008,dobbs2010}.~~\cite{heng2012} investigated supersonic flows (flows that approach or exceed $\approx 1$~kms$^{-1}$) in irradiated exoplanetary atmospheres and found that shocks form on the dayside hemisphere upstream of the substellar point due to the enhanced temperatures that develop there.

Modelling of the atmospheric circulation of the hot neptune GJ436b has given insight into the effect of metallicity on the dynamical and radiative behaviour of extrasolar giant planetary atmospheres.  As the atmospheric metallicity increase from solar levels to $50\times$ solar, the maximum zonal wind speeds reached increase from $\approx1.3$~kms$^{-1}$ to over $2$~kms$^{-1}$, suggesting that  zonal wind speeds increase with increasing metallicity~\citep{lewis2010}. 

The meteorology of solar system planets has also been extensively studied:  tracked cloud features on Uranus~\citep{hammel2001,sromovsky2005} and Saturn~\citep{porco2005} imply wind speeds of $\approx O(0.1$~kms$^{-1})$, which are too slow for Alfv\'{e}n ionization to be a feasible process in their atmospheres.  However, this refers only to the observable part of Jovian atmosphere's.

Our aim here is not to emulate such seminal meteorological models; but instead to examine simple atmospheric balanced flows to construct a compelling and feasible argument for the attainment of the required $v_{c}$ for Alfv\'{e}n ionization.  Balanced flows are an idealized view of actual atmospheric motion, but often give a good approximation of actual flows and yield a good estimation of the characteristic flow speeds.  On a fundamental level, meteorological phenomena are driven by pressure variations and rotation and so it is possible to cast $v_{c}$ in terms of the required pressure gradient to drive neutral gas flows for a given rotation frequency.  The underlying physical mechanism that drive the pressure variations will vary depending on the specific object in question.  For example, for planetary bodies differential heating caused by illumination from a stellar companion will be the primary driver; but for substellar objects magnetic activity and/or resistive heating may be more significant.  We will investigate this in a forthcoming paper.  For simplicity, in the following treatment we do not prescribe a specific process but look at the consequent physical ramifications of the atmospheric pressure variations produced by such processes.  Here we look at three standard balanced flows:  geostrophic, cyclostrophic and gradient flows.  For a more detailed discussion regarding these balanced flows see appendix~\ref{appendix}.

\subsubsection{Geostrophic flow}

Consider a local cartesian tangent plane $(x,y)$ with latitude $\phi$ on the surface of a spherical body, rotating with angular speed $\omega$.  In a rotating reference frame a fluid element of atmospheric gas experiences the pressure gradient force, the Coriolis effect, the centrifugal effect and gravity.  Geostrophic wind balance results from the balance between the Coriolis effect and the pressure gradient force,  
\begin{equation}
u_{y}=\frac{1}{f\rho_{\rm gas}}\frac{\partial p_{\rm gas}}{\partial x},
\end{equation}
where $f=2\omega\sin{\phi}$ is the Coriolis parameter characterising the significance of the Coriolis effect as a function of position on the surface.  For Alfv\'{e}n ionization to be a feasible process we require gas flows of $|u_{y}|\approx O(10~\textnormal{kms}^{-1})$.  This places the following constraint on the pressure gradient required to drive such flows,
\begin{equation}
\left|\frac{\partial p_{\rm gas}}{\partial x}\right|\geq f\rho_{\rm gas}~~[\textnormal{bar~cm}^{-1}], \label{geo}
\end{equation}
where $f$ and $\rho_{\rm gas}$ are in [rad s$^{-1}$] and [g~cm$^{-3}$] respectively.  Note that the magnitude of the pressure gradient is proportional to the gas mass density $\rho_{\rm gas}$.  Geostrophic flows are a good approximation to actual wind flows outside of the tropics (extratropical)~\citep{holton2004}

\subsubsection{Cyclostrophic flow}
Transforming from cartesian to cylindrical coordinates $(r,\theta,z)$, in a locally rotating non-inertial frame (such as a storm system or cyclone) an atmospheric fluid element experiences local centrifugal and local Coriolis effects in addition to the global effects of the rotating body.  Cyclostrophic flows are characterized by the force balance between the pressure gradient force and the local centrifugal effect,
\begin{equation}
u_{\theta}=\left(\frac{r}{\rho_{\rm gas}}\frac{\partial p_{\rm gas}}{\partial r}\right)^{1/2}.
\end{equation}
For Alfv\'{e}n ionization ($|u_{\theta}|\approx O(10~\textnormal{kms}^{-1})$), this is achievable if the pressure gradient obeys
\begin{equation}
\left|\frac{\partial p_{\rm gas}}{\partial r}\right|\geq1\times10^{6}\frac{\rho_{\rm gas}}{r}~~[\textnormal{bar~cm}^{-1}] ,\label{cyclo}
\end{equation}
where $r$ and $\rho_{\rm gas}$ are in [cm] and [g~cm$^{-3}$] respectively.  As in the geostrophic case the required pressure gradient is proportional to the gas mass density $\rho_{\rm gas}$ and is additionally inversely proportional to the radius of the weather system $r$.  Therefore smaller (larger) storm systems will yield severer (more lenient) pressure gradient criterion.  Please note that in the subsequent analysis we shall consider a weather system with circular motion of radius $r\approx1000$~km.  This is comparable to the white oval storm systems observed on Jupiter ~\citep{beebe1997,choi2010}.  The Great Red Spot storm system has a spatial extent of the order $O(10^4~\textnormal{km})$~\citep{beebe1997}, and so greater storm systems are possible.  Cyclostrophic flows are characteristic of weather systems such as tornadoes~\citep{holton2004}. 

In comparison, geostrophic and cyclostrophic flows only differ by a multiplicative factor, depending on the angular frequency and the radius of the weather system: 
\begin{equation}
\frac{|\partial p_{\rm gas}/\partial x|}{|\partial p_{\rm gas}/\partial r|}=1\times10^{-6}fr.
\end{equation}

\subsubsection{Gradient flow}
Remaining in cylindrical coordinates, gradient flows are characterized by the balance between the global Coriolis effect, the pressure gradient force and the local centrifugal effect,
\begin{equation}
u_{\theta}=-\frac{1}{2}rf\pm\frac{1}{2}r\left(f^{2}+\frac{4}{r\rho_{\rm gas}}\frac{\partial p_{\rm gas}}{\partial r} \right)^{1/2}.
\end{equation}
For flow speeds $|u_{\theta}|\approx O(10~\textnormal{kms}^{-1})$, this will occur if the pressure gradient obeys,
\begin{equation}
\left|\frac{\partial p_{\rm gas}}{\partial r}\right|\geq2.5\times10^{-7}r\rho_{\rm gas}\left[ \frac{1}{r^{2}}(2\times10^{6}-rf)^{2}-f^{2}\right] ~~[\textnormal{bar~cm}^{-1}].\label{grad}
\end{equation}
where $r$ and $\rho_{\rm gas}$ are in [cm] and [g~cm$^{-3}$] respectively.  Note that the pressure gradient criterion is directly proportional to the gas mass density $\rho_{\rm gas}$; and that for a given value of $r$ and $f$, it is equal to a constant times $\rho_{\rm gas}$.  With increasing rotation period, $f\rightarrow 0$ and gradient flows reduce to cyclostrophic flows.  In the following analysis we will consider a low-pressure system ($\partial p_{\rm gas}/\partial r>0$ and $u_{\theta}>0$) with circular motion of radius $r\approx1000$~km.  

In conclusion, for the balanced flows considered here, the pressure gradient criteria are proportional to $\rho_{\rm gas}$, for a given rotation period and/or weather system radius (characterized by the Coriolis parameter $f$ and radius $r$ respectively).  Therefore, depending on the chosen values of $f$ and $r$, the relative performance between the three flows can be different.  For example, for a storm system with a small (large) radius $r$, cyclostrophic flows could produce a pressure gradient criterion that has a greater (smaller) magnitude than that of geostrophic flows.

\subsubsection{Flows in substellar atmospheres}

Consider a low-mass object rotating with angular frequency $\omega=3.5\times10^{-4}$~rads$^{-1}$ (i.e. a rotation period of $T=5$~hrs) therefore the Coriolis parameter is $f\approx4.9\times10^{-4}$~rads$^{-1}$ for $\phi=\pi/4$ (mid-latitudes).  Exoplanets are expected to have rotation periods ranging from $1-4$~days for Hot Jupiters; $10-20$~hrs for Jupiter-like planets; and $2-3$~hrs for fast-rotators approaching centrifugal breakup~\citep{sanchez2001,sanchez2004}.  Brown dwarfs are considered to be very rapid rotators with average rotation periods $\leq 1$~day, with a lower limit of $\approx2-4$~hrs~\citep{scholz2005,joergens2003,osorio2006}.  M-dwarfs have been observed to have rotation periods ranging from $\approx 7-150$~hrs~\citep{reiners2012b}.  Radio activity (and hence the presence of an atmosphere plasma) appears to be correlated with fast rotators justifying our choice of rotation period $T=5$~hrs, which is close to the centrifugal breakup of the low-mass objects considered here.

\begin{figure}
\includegraphics{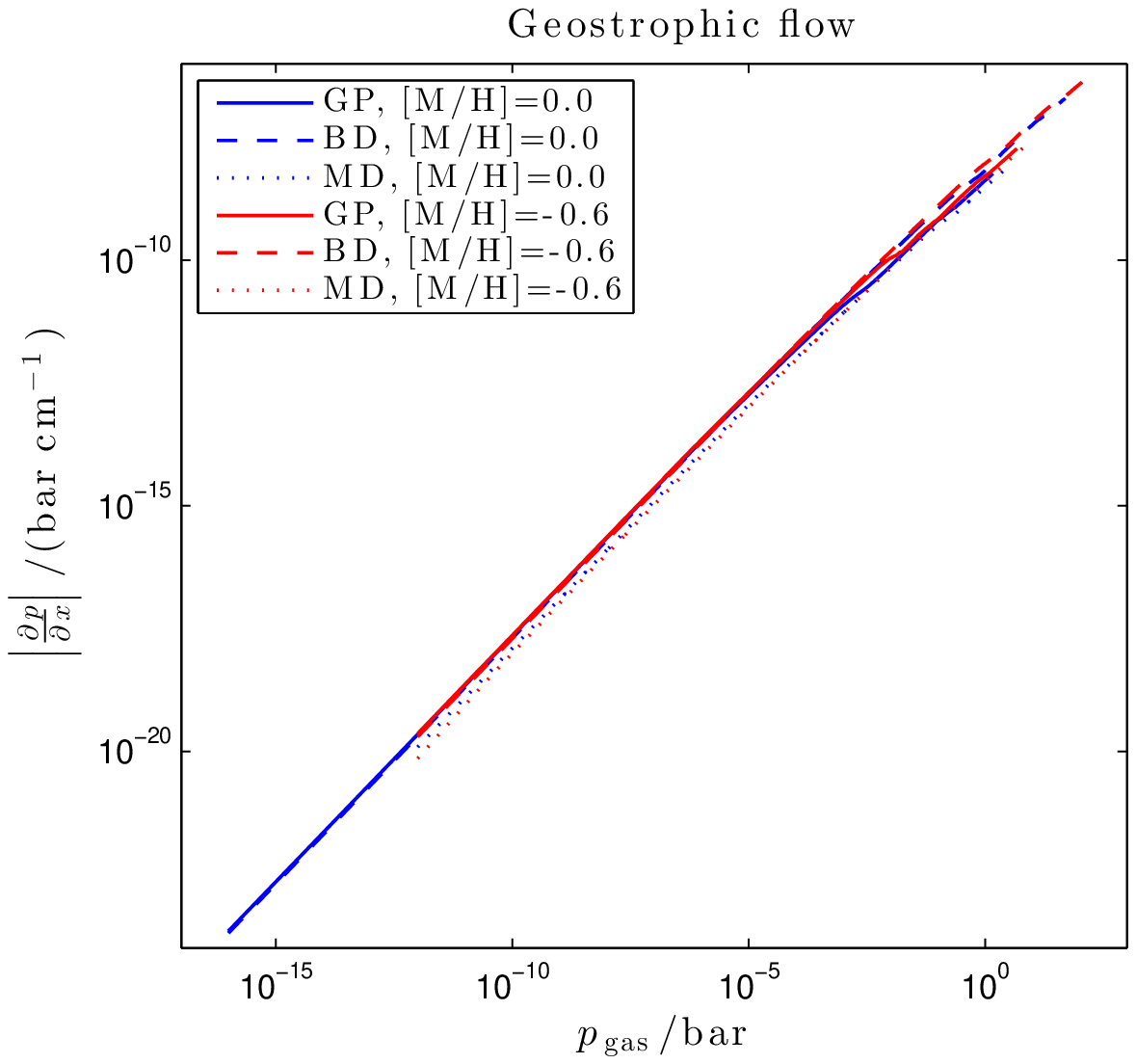}
\caption{The pressure gradient criterion for a geostrophic flow (Eqn.~\ref{geo}) for all three groups of low-mass object considered here, with solar elemental abundances and low metallicity.  This plot shows the minimum required pressure gradient needed to drive flows at the critical speed ($|u_{y}|\approx O(10~\textnormal{kms}^{-1})$) needed for Alfv\'{e}n ionization.  We have assumed a typical low-mass object with rotation angular frequency $\omega=3.5\times10^{-4}$~rads$^{-1}$ ($T=5$~hrs) the Coriolis parameter is $f\approx4.9\times10^{-4}$~rads$^{-1}$ for $\phi=\pi/4$ (mid-latitudes). \label{dp_geo}}
\end{figure}

\begin{figure}
\includegraphics{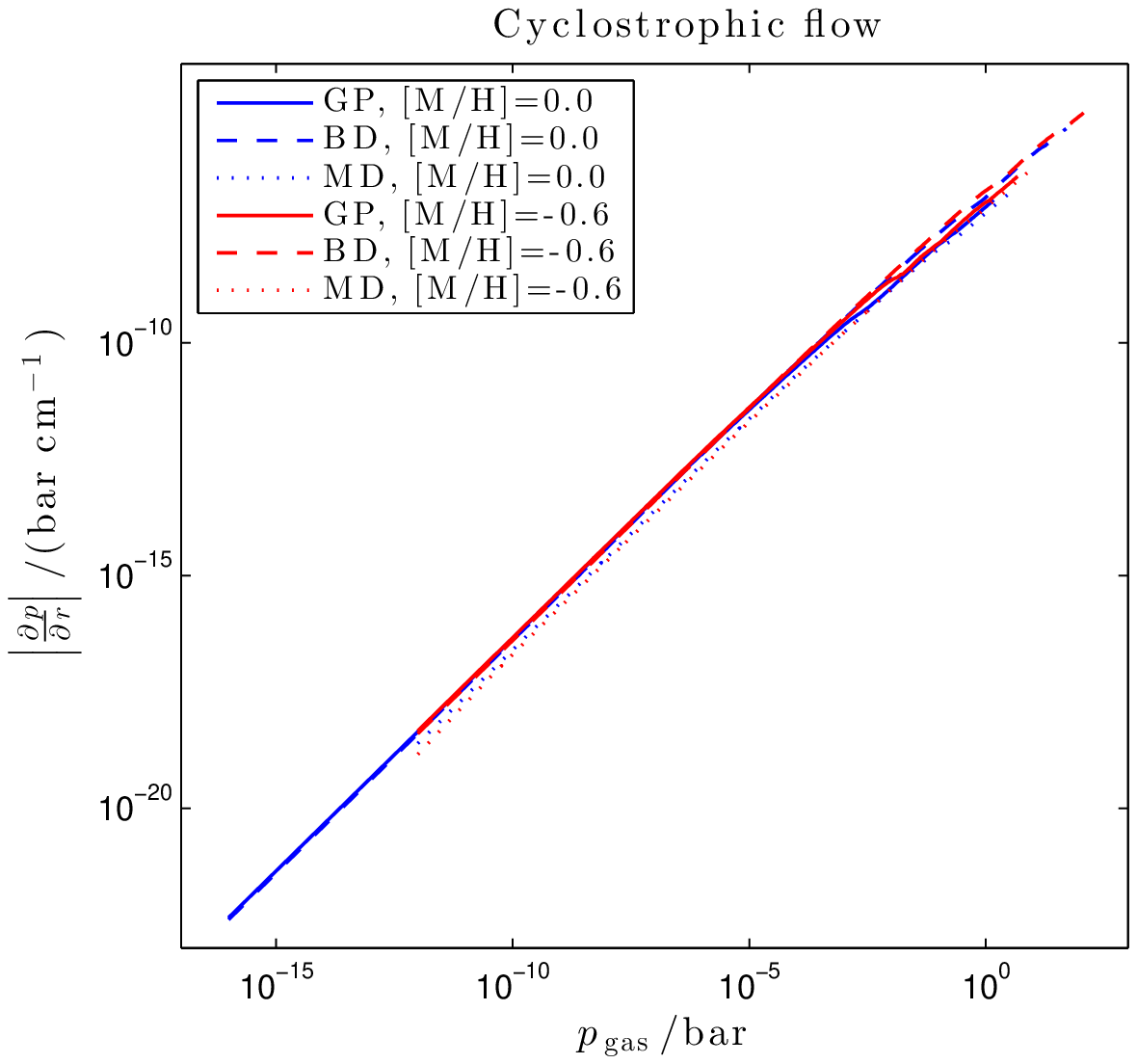}
\caption{The pressure gradient criterion for a cyclostrophic flow (Eqn.~\ref{cyclo}) for all three groups of low-mass object considered here, with solar elemental abundances and low metallicity.  This plot shows the minimum required pressure gradient needed to drive flows at the critical speed ($|u_{y}|\approx O(10~\textnormal{kms}^{-1})$) needed for Alfv\'{e}n ionization.  We have assumed a typical low-mass object with rotation angular frequency $\omega=3.5\times10^{-4}$~rads$^{-1}$ ($T=5$~hrs) the Coriolis parameter is $f\approx4.9\times10^{-4}$~rads$^{-1}$ for $\phi=\pi/4$ (mid-latitudes). \label{dp_cyclo}}
\end{figure}

\begin{figure}
\includegraphics{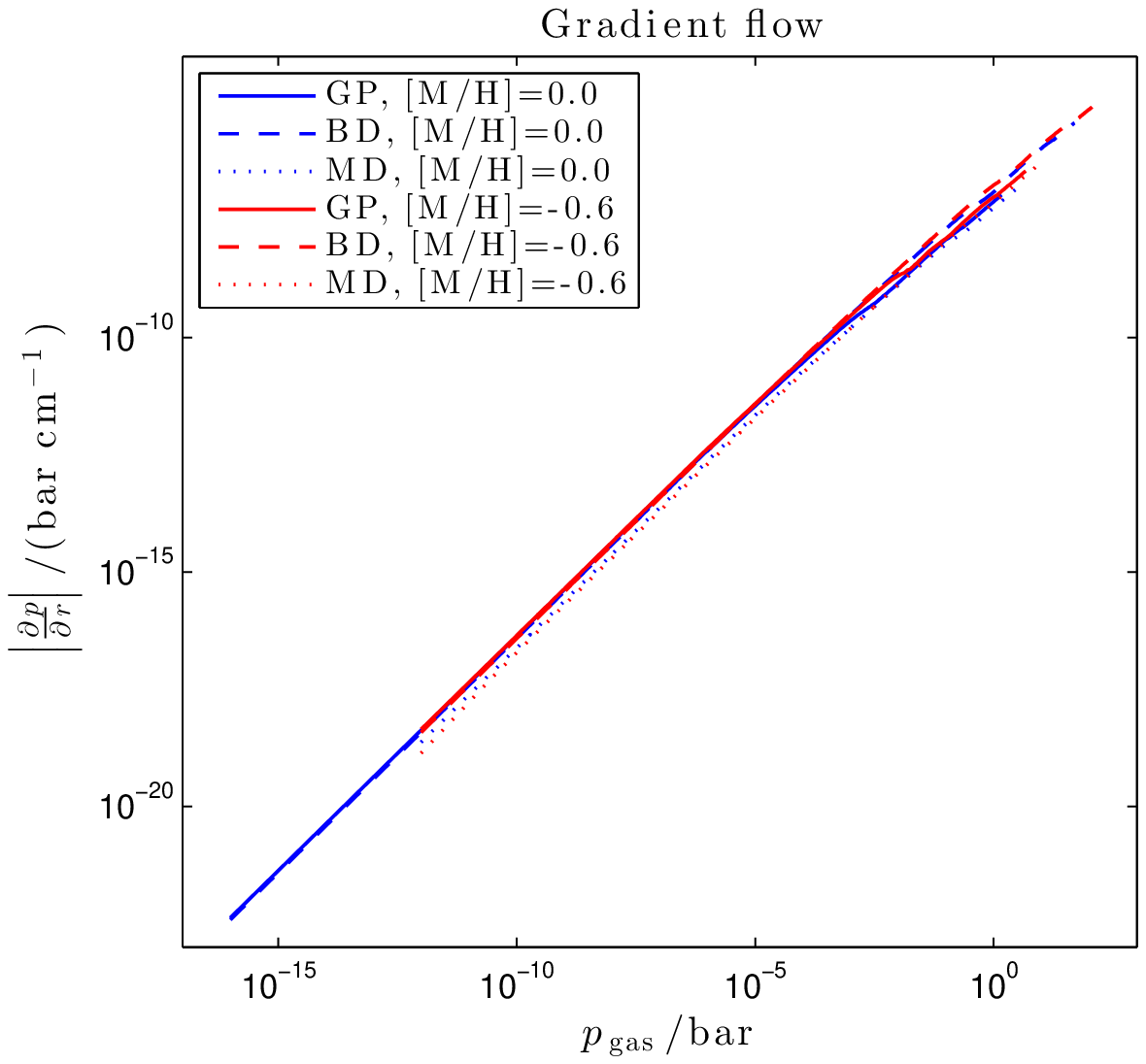}
\caption{The pressure gradient criterion for a gradient flow (Eqn.~\ref{grad}) for all three groups of low-mass object considered here, with solar elemental abundances and low metallicity.  This plot shows the minimum required pressure gradient needed to drive flows at the critical speed ($|u_{y}|\approx O(10~\textnormal{kms}^{-1})$) needed for Alfv\'{e}n ionization.  We have assumed a typical low-mass object with rotation angular frequency $\omega=3.5\times10^{-4}$~rads$^{-1}$ ($T=5$~hrs) the Coriolis parameter is $f\approx4.9\times10^{-4}$~rads$^{-1}$ for $\phi=\pi/4$ (mid-latitudes). \label{dp_grad}}
\end{figure}

Figs.~\ref{dp_geo},~\ref{dp_cyclo} and~\ref{dp_grad} show the pressure gradient criteria for geostrophic (Eqn.~\ref{geo});  cylcostrophic  (Eqn.~\ref{cyclo}); and gradient flows (Eqn.~\ref{grad}) respectively, exhibiting how they vary throughout the atmosphere $(p_{\rm gas},T_{\rm gas})$ for all three groups of low-mass object considered here, with solar elemental abundances and low metallicity.   These plots show the minimum required pressure gradient needed to drive flows at the critical speed ($|u_{\theta}|\approx O(10~\textnormal{kms}^{-1})$) needed for Alfv\'{e}n ionization to occur.  {\sc Drift-Phoenix} models obey 1D hydrostatic equilibrium, hence no horizontal wind velocity profiles can be provided.  

For all the balanced flows considered here, the pressure gradient criteria are equal to a multiplicative factor times the gas mass density $\rho_{\rm gas}$ (for a given $f$ and $r$).  Therefore, for a single model atmosphere, the different balanced flows will have the same functional form and will only differ by a multiplicative factor.  For example, for a particular model atmosphere, the gradient and cyclostrophic criteria for $f\approx4.9\times10^{-4}$~rads$^{-1}$ and $r=1000$~km are indistinguishable (Figs.~\ref{dp_cyclo}~and~\ref{dp_grad}).  Moreover, since the different model atmospheres considered here have similar $(p_{\rm gas},\rho_{\rm gas})$ profiles the respective pressure gradient criteria are similar.  However, GP and BD models for [M/H]$=0.0$ extend to lower mass densities and pressures in comparison to the other models considered here (Fig.~\ref{t_p_m}) and so in these atmospheres where the pressure is low, it is easier to fulfil the criteria for the balanced flows discussed here.

In general, the pressure gradient is directly proportional to the gas density $\rho_{\rm gas}$ and so Figs.~\ref{dp_geo},~\ref{dp_cyclo} and~\ref{dp_grad} show that at high altitudes ($p_{\rm gas}\approx 10^{-16}-10^{-12}$~bar) where the atmosphere is diffuse and the density is low $\rho_{\rm gas}\approx10^{-20}-10^{-17}$~g~cm$^{-3}$, then the required pressure gradient to drive the needed flows is $|\partial p_{\rm gas}/\partial x|\gtrsim (10^{-25}-10^{-18})$~bar~cm$^{-1}$.  At lower altitudes ($p_{\rm gas}\approx 1-10^{2}$~bar) the atmospheric density is clearly higher ($\rho_{\rm gas}\approx10^{-5}-10^{-3}$~g~cm$^{-3}$) and so to obtain similar flows the required pressure gradient is much larger $|\partial p_{\rm gas}/\partial x|\approx10^{-8}-10^{-6}$~bar~cm$^{-1}$.  Typically for the Earth pressure gradients are of the order $|\partial p_{\rm gas}/\partial x|\approx10^{-7}$~bar~cm$^{-1}$ in the troposphere or at mid-latitudes $\approx10^{-9}$~bar~cm$^{-1}$, with much greater pressure gradients expected at fronts.  With this in mind it seems plausible that in more exotic environments, namely the atmosphere of a low-mass object, such pressure gradients, and hence wind speeds, required for Alfv\'{e}n ionization seem achievable.

Figs.~\ref{dp_geo},~\ref{dp_cyclo} and~\ref{dp_grad} show that at low atmospheric pressures the required pressure gradient criteria is smaller than at higher pressures; this conveniently coincides with the region of the atmosphere where the criteria for a magnetized seed plasma is easiest to achieve (Fig.~\ref{b_p}). 

\begin{figure}
\includegraphics{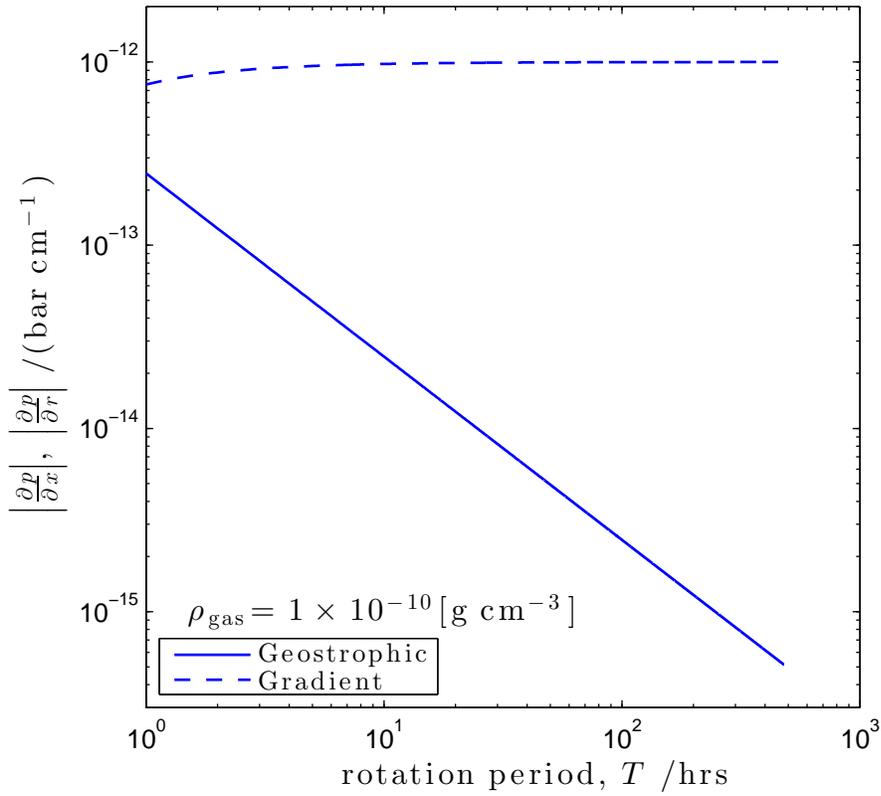}
\caption{Pressure gradient as a function of rotational period $T$ for a geostrophic flow (Eqn.~\ref{geo}, solid line) and a gradient flow (Eqn.~\ref{grad}, dashed line).  This is the critical pressure gradient required to provide large enough local gas flow speeds to allow Aflv\'{e}n ionization. \label{dp_t}}
\end{figure}

Figs.~\ref{dp_geo},~\ref{dp_cyclo} and~\ref{dp_grad} show the dependence of the different flow regimes on the atmospheric density where a low-mass object rotating with a constant angular frequency $\omega=3.5\times10^{-4}$~rads$^{-1}$ ($T=5$~hrs and $f\approx4.9\times10^{-4}$~rads$^{-1}$ for $\phi=\pi/4$) is considered.  However, geostrophic and gradient flow speeds also depend on the rotation period of the object (and hence the Coriolis paramter), Fig.~\ref{dp_t} is a plot of Eqns.~\ref{geo} and~\ref{grad} as a function of rotation period $T$ for a constant atmospheric density.  Without loss of generality, we have chosen $\rho_{\rm gas}\approx10^{-10}$~g~cm$^{-3}$  since it is typical value for the density at mid-atmospheric pressures ($p_{\rm gas}\approx10^{-5}$~bar) for the model atmospheres studied here (Fig.~\ref{t_p_m}).  The pressure gradient criterion for geostrophic flows is linearly dependent on the Coriolis parameter $f=2\omega\sin{\phi}$ and therefore inversely proportional to the period $T$ (Eqn.~\ref{geo}).  Thus, Fig.~\ref{dp_t} shows that for geostrophic flows as the rotation period of the low-mass object increases the required pressure gradient for Alfv\'{e}n ionization to occur decreases. Therefore it is relatively easier for slower rotating objects to meet the flow speed criterion.  For example, an object with rotation period $T\approx2$~hrs, the pressure gradient must exceed $\approx10^{-9}$~bar~cm$^{-1}$; whereas for $T\approx10^{2}$~hrs, $|\partial p_{\rm gas}/\partial x|\gtrsim10^{-11}$~bar~cm$^{-1}$.  For gradient flows, as $T$ increases so does the required minimum pressure gradient until it saturates at a maximum value $|\partial p_{\rm gas}/\partial r|\gtrsim10^{-12}$~bar~cm$^{-1}$.  In the limit of large $T$ (i.e. $f\rightarrow0$), Eqn.~\ref{grad} becomes $|\partial p_{\rm gas}/\partial r|\gtrsim10^{6}\rho_{\rm gas}/r$; therefore, for $r\approx1000$~km, $|\partial p_{\rm gas}/\partial r|\gtrsim10^{-12}$~bar~cm$^{-1}$.
\section{Degree of ionization from Alfv\'{e}n ionization\label{degree}}

For Alfv\'{e}n ionized we require that: (i) the seed plasma is strongly magnetized; and (ii) the neutral gas flow reaches a critical speed $v_{c}$.  Assuming the required conditions can be met and the neutral gas flow speed can be maintained, Alfv\'{e}n ionization can ionize the entirety of the gas in a localized volume, leaving a plasma with an electron number density equal to the gas component number density (assuming 100\% ionization) plus the initial seed magnetized plasma number density.  This is assuming the gas is singly ionized.  100\% ionization via the Alfv\'{e}n mechanism is perfectly feasible and has been demonstrated for Hydrogen in laboratory experiments~\citep{fahleson1961}.  Here, we are interested in Alfv\'{e}n ionization in the atmospheres of gas giant planets, brown dwarfs and M-dwarfs.  Therefore, we assume the seed plasma is sourced from the ions and electrons created by thermal ionization in the atmosphere and we exclude any contributions from additional sources such as inter-grain discharges and cosmic-rays.  In a gas mixture composed of a number of species, the electron number density can't exceed the target species neutral number density plus the initial seed electron number density.  For a given target species this can enhance the local degree of ionization $f_{e}=n_{i}/(n_{i}+n_{gas})$ to beyond $10^{-7}$, resulting in a ionized gas that can be treated as a plasma.  

\begin{figure}
\includegraphics{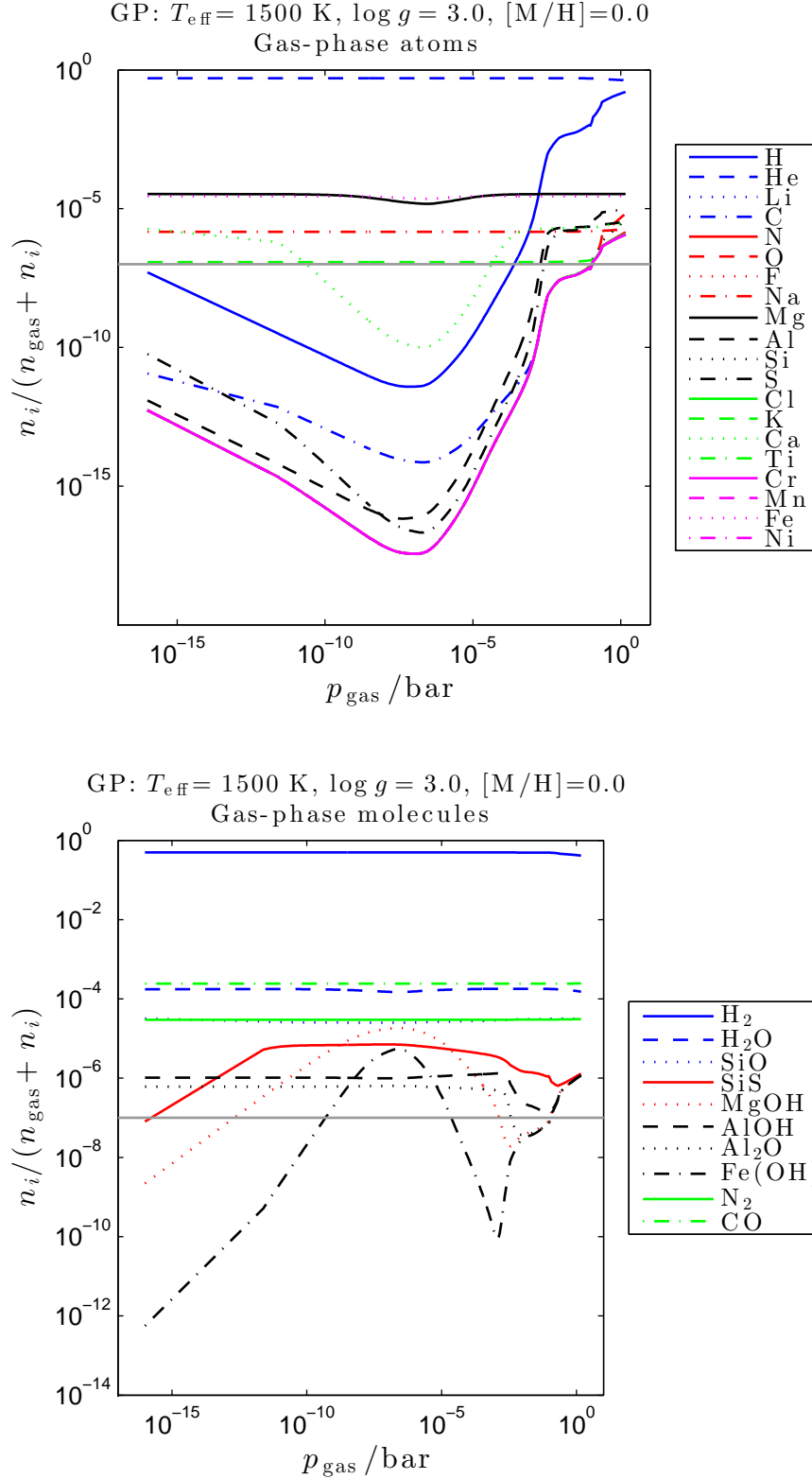}
\caption{The degree of ionization $f_{e}=n_{i}/(n_{\rm gas}+n_{i})$ resulting from Alfv\'{e}n ionization for a gas giant planet (GP: $T_{\rm eff}=1500$~K, $\log{g}=3.0$) for individual species being entirely ionized in a localized atmospheric pocket, assuming initially solar elemental abundances ([M/H] $=0.0$).  The top plot shows $f_{e}$ for atoms; and the bottom plot shows $f_{e}$ for selected molecules.  The grey line signifies $f_{e}=10^{-7}$, the degree of ionization required to constitute a plasma.  \label{gp_s_1}}
\end{figure}

\begin{figure}
\includegraphics{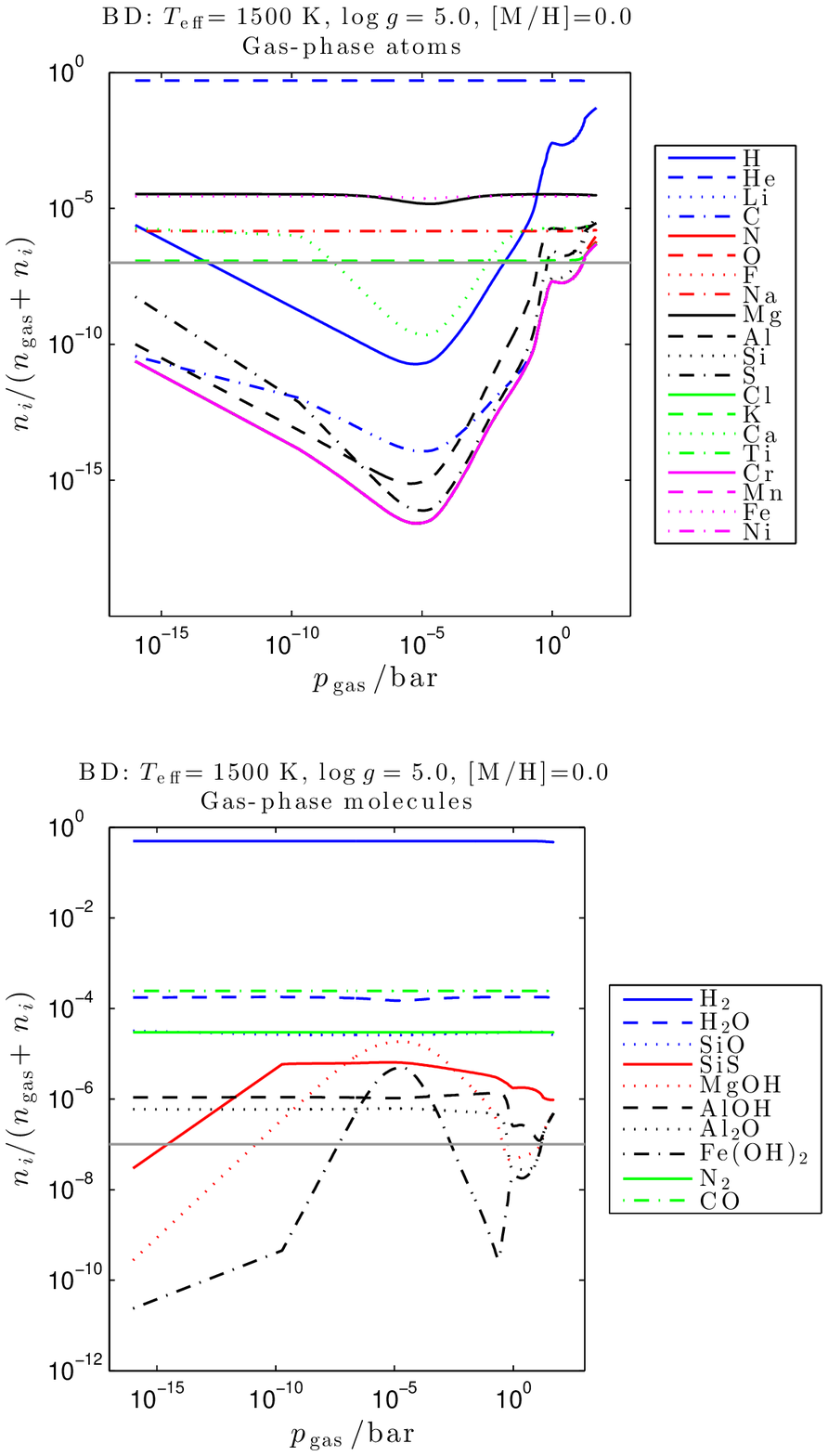}
\caption{The degree of ionization $f_{e}=n_{i}/(n_{\rm gas}+n_{i})$ resulting from Alfv\'{e}n ionization for a brown dwarf (BD: $T_{\rm eff}=1500$~K, $\log{g}=5.0$) for individual species being entirely ionized in a localized atmospheric pocket, assuming initially solar elemental abundances ([M/H] $=0.0$). The top plot shows $f_{e}$ for atoms; and the bottom plot shows $f_{e}$ for selected molecules.  The grey line signifies $f_{e}=10^{-7}$, the degree of ionization required to constitute a plasma.  \label{bd_s_1}}
\end{figure}

\begin{figure}
\includegraphics{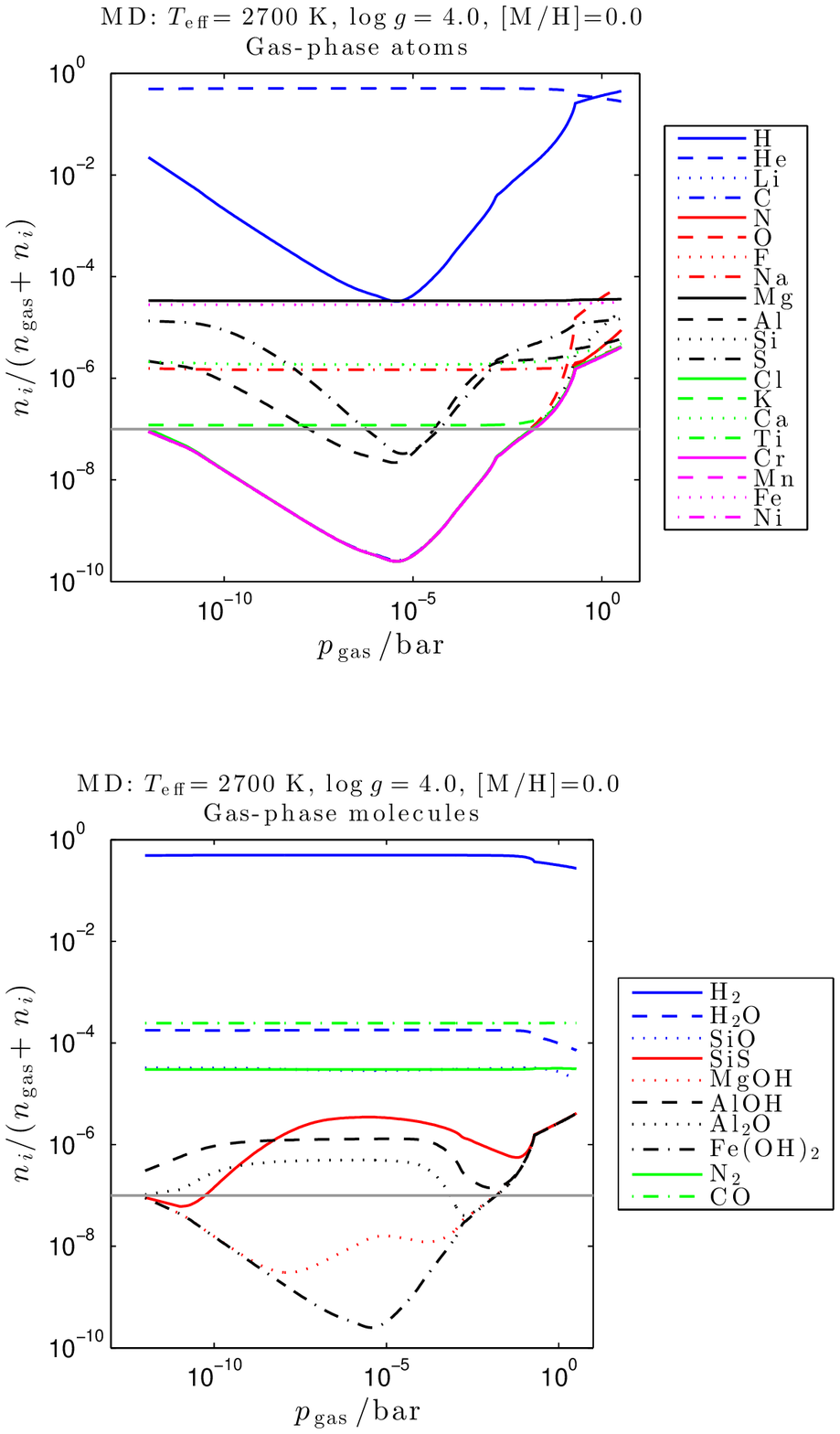}
\caption{The degree of ionization $f_{e}=n_{i}/(n_{\rm gas}+n_{i})$ resulting from Alfv\'{e}n ionization for a M-dwarf (MD: $T_{\rm eff}=2700$~K, $\log{g}=4.0$) for individual species being entirely ionized in a localized atmospheric pocket, assuming initially solar elemental abundances ([M/H] $=0.0$). The top plot shows $f_{e}$ for atoms; and the bottom plot shows $f_{e}$ for selected molecules.  The grey line signifies $f_{e}=10^{-7}$, the degree of ionization required to constitute a plasma.   \label{md_s_1}}
\end{figure}

\begin{figure}
\includegraphics{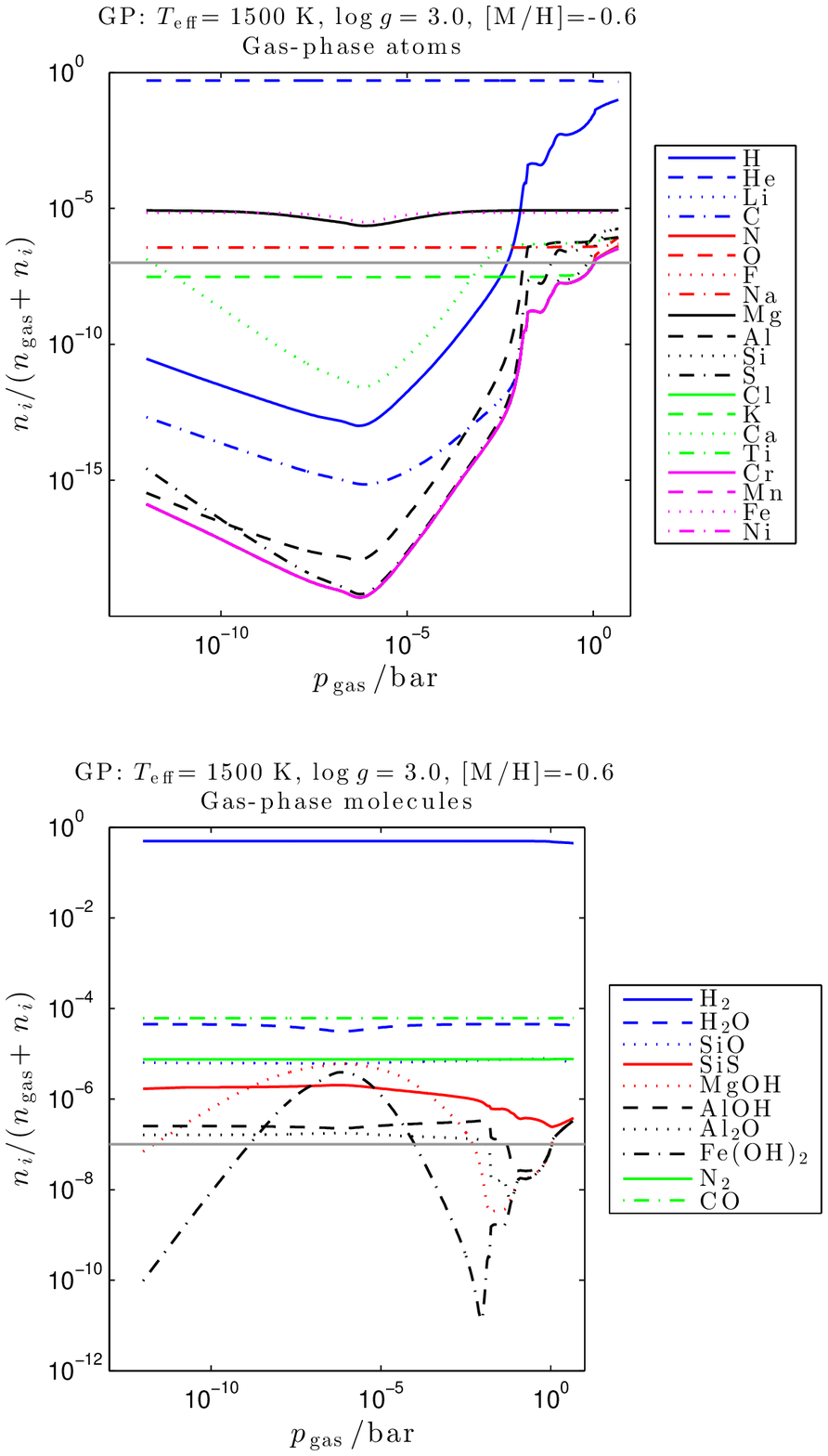}
\caption{The degree of ionization $f_{e}=n_{i}/(n_{\rm gas}+n_{i})$ resulting from Alfv\'{e}n ionization for a gas giant planet (GP: $T_{\rm eff}=1500$~K, $\log{g}=3.0$) for individual species being entirely ionized in a localized atmospheric pocket, assuming initially low metallicity ([M/H] $=-0.6$). The top plot shows $f_{e}$ for atoms; and the bottom plot shows $f_{e}$ for selected molecules.  The grey line signifies $f_{e}=10^{-7}$, the degree of ionization required to constitute a plasma.  \label{gp_m_1}}
\end{figure}

\begin{figure}
\includegraphics{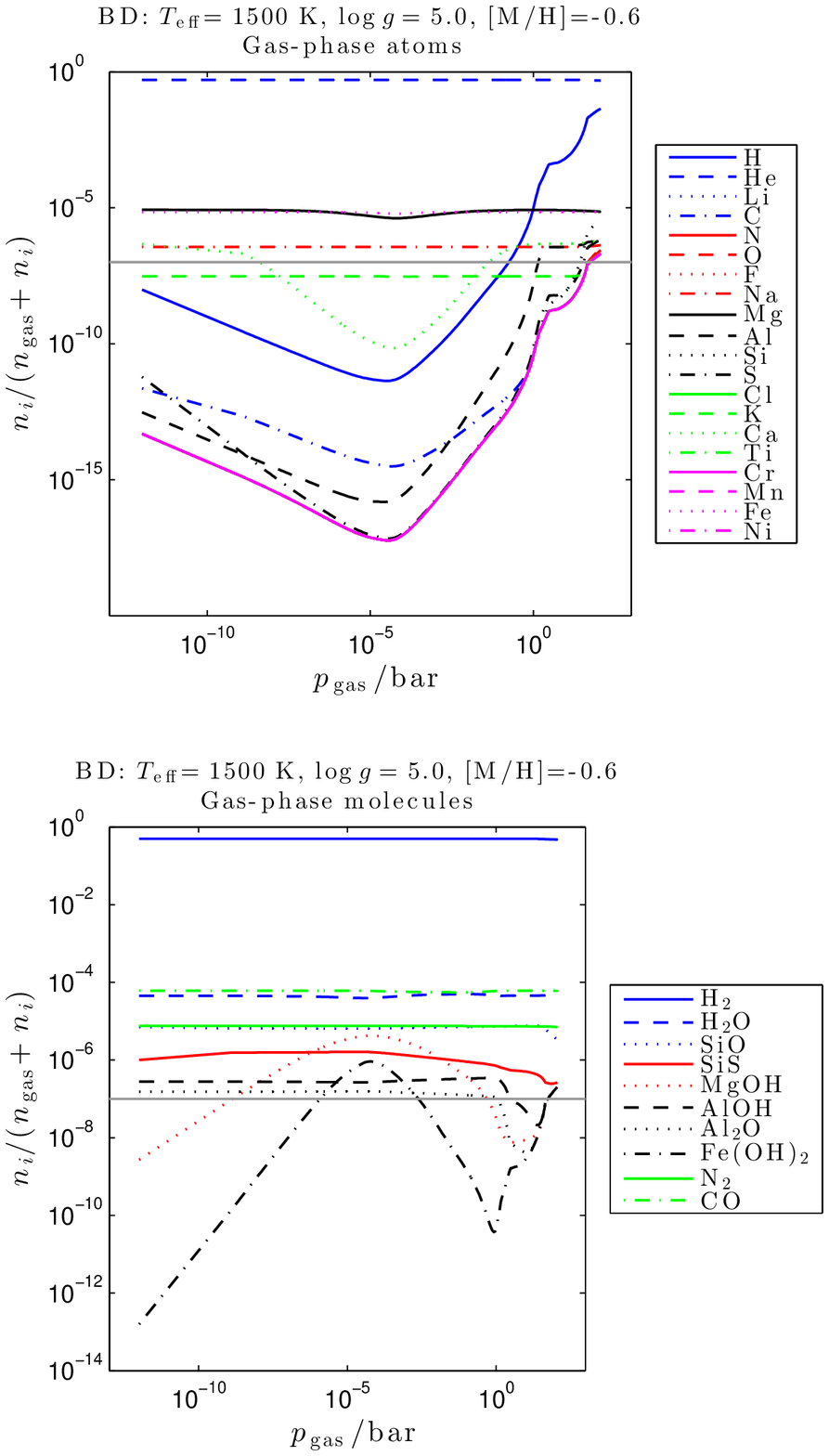}
\caption{The degree of ionization $f_{e}=n_{i}/(n_{\rm gas}+n_{i})$ resulting from Alfv\'{e}n ionization for a brown dwarf (BD: $T_{\rm eff}=1500$~K, $\log{g}=5.0$) for individual species being entirely ionized in a localized atmospheric pocket, assuming initially low metallicity ([M/H] $=-0.6$). The top plot shows $f_{e}$ for atoms; and the bottom plot shows $f_{e}$ for selected molecules.  The grey line signifies $f_{e}=10^{-7}$, the degree of ionization required to constitute a plasma.\label{bd_m_1}}
\end{figure}

\begin{figure}
\includegraphics{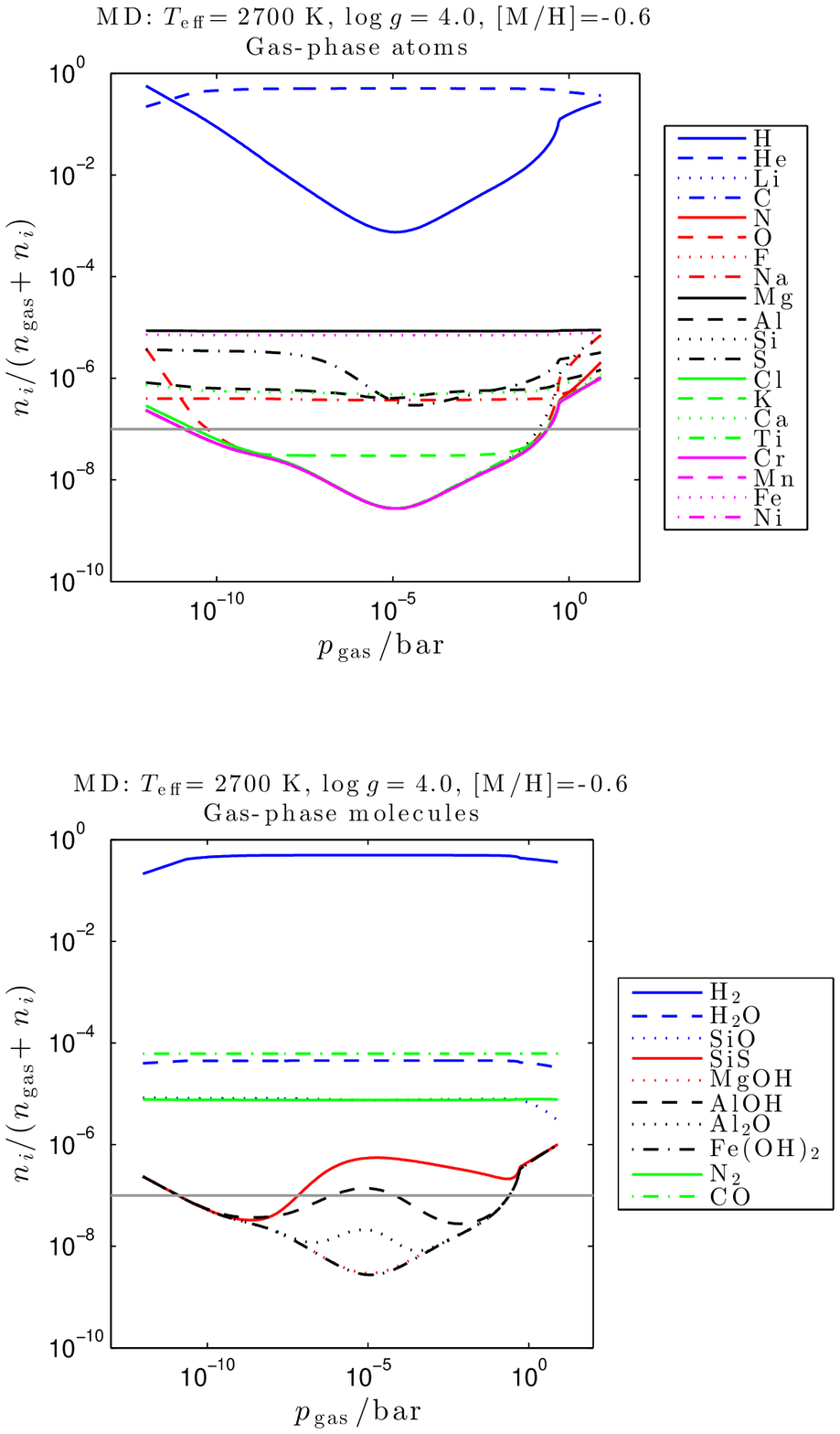}
\caption{The degree of ionization $f_{e}=n_{i}/(n_{\rm gas}+n_{i})$ resulting from Alfv\'{e}n ionization for a M-dwarf (MD: $T_{\rm eff}=2700$~K, $\log{g}=4.0$) for individual species being entirely ionized in a localized atmospheric pocket, assuming initially low metallicity ([M/H] $=-0.6$). The top plot shows $f_{e}$ for atoms; and the bottom plot shows $f_{e}$ for selected molecules.  The grey line signifies $f_{e}=10^{-7}$, the degree of ionization required to constitute a plasma.\label{md_m_1}}
\end{figure}

Figs.~\ref{gp_s_1}--\ref{md_m_1} show the resulting degrees of ionization from Alfv\'{e}n ionization,  if specific individual species constituting the atmospheric gas are entirely ionized (on their own) in a localized atmospheric pocket, assuming solar elemental abundances [M/H]~$=0.0$ (Figs.~\ref{gp_s_1}--\ref{md_s_1}) and a lower metallicity of [M/H]~$=-0.6$ (Figs.~\ref{gp_m_1}--\ref{md_m_1}).   In general,  if in a localized atmospheric pocket a particular species can be $100\%$ ionized, then the species with the greatest number density will yield the highest degree of ionization.  To summarise: if entirely ionized on their own He, Fe, Mg, Na, H$_{2}$, CO, H$_{2}$O, N$_{2}$ and SiO all consistently increase the degree of ionization beyond $10^{-7}$ throughout the model atmospheres considered here.  

In all the atmosphere models studied here He, Fe, Mg and Na, if individually ionized (100\%) in a localized atmospheric volume, increase the degree of ionization sufficiently high ($f_{e}\approx1,~10^{-5},~10^{-5}$ and~$10^{-6}$~respectively) that the ionized gas can be considered a plasma ($f_{e}\geq10^{-7}$)  throughout the atmosphere.  The critical ionization speed that a streaming neutral gas must obtain for the ionization of these respective atoms is $v_{c}=34.43$~kms$^{-1}$,~$5.22$~kms$^{-1}$, $7.79$~kms$^{-1}$ and $6.57$~kms$^{-1}$.  He, Fe, Mg and Na have the greatest number density of the atomic chemical elements in the atmospheres.  In addition Mg and Fe play a key role in the formation of dust clouds which is evident from a dip in their atmospheric $(p_{\rm gas},T_{\rm gas})$ profile in the cloud forming region (Fig.~\ref{t_p_m}).  Helium is inert and so is not affected by the formation of clouds and its number density doesn't vary significantly from the initial, prescribe abundance.  However, although chemically inert, Helium metastable states can affect the energy budget of the system under consideration by momentarily locking energy in an excited state.  Note that the number density of atomic He is greater than that of H; molecular Hydrogen H$_{2}$ is the dominant H-carrier, the formation of which depletes the atomic H in all the atmospheres considered here for $p_{\rm gas}\lesssim 10^{-1}$~bar.  

The low-metallicity model atmospheres ([M/H]$~=-0.6$, Figs.~\ref{gp_m_1},~\ref{bd_m_1} and~\ref{md_m_1}) have a lower number density of metals relative to hydrogen, and so the degree of ionization from these metals is lower than in the solar metallicity case since their number density is reduced.  Potassium ($v_{c}=4.22$~kms$^{-1}$) can also be significant in enhancing the degree of ionization to form an atmospheric plasma, but only in solar abundance models (Figs.~\ref{gp_s_1},~\ref{bd_s_1}~and~\ref{md_s_1});  in the low metallicity models the degree of ionization falls below the threshold value of $10^{-7}$ over the majority of the atmosphere, only succeeding to exceed $10^{-7}$ at very high pressures (in all three groups of low-mass object considered here) and very low pressures (MD  only, Fig.~\ref{md_s_1} and~\ref{md_m_1}).  Calcium ($v_{c}=5.42$~kms$^{-1}$) amplifies the degree of ionization beyond $10^{-7}$ for all pressures in MD atmospheres (even for a low metallicity atmosphere) but only for high and low pressures in GP (Fig.~\ref{gp_s_1}~and~\ref{gp_m_1}) and BD atmospheres (Fig.~\ref{bd_s_1}~and~\ref{bd_m_1}), which are considerably cooler than the M-dwarf model atmosphere. For the remaining elements, their relative number density in the gas-phase is so small that if entirely ionized they, on their own, only contribute significantly to the generation of a plasma at high atmospheric pressures.  

Whereas the GP (Fig.~\ref{gp_s_1}~and~\ref{gp_m_1}) and BD (Fig.~\ref{bd_s_1}~and~\ref{bd_m_1}) models have a similar $f_{e}$ behaviour, the MD model (Fig.~\ref{md_s_1}~and~\ref{md_m_1}) has significant differences.  For example, after helium, hydrogen yields the highest degree of ionization ($f_{e}\approx10^{-4}$ at $p_{\rm gas}\approx10^{-5}$) which is not the case in the GP and BD atmospheric models.  Moreover, in contrast to the GP and BD models; calcium, aluminium and sulphur yield a degree of ionization greater than $10^{-7}$ for most atmospheric pressures.  In low metallicity MD atmospheres (MD, [M/H]$~=-0.6$), additional elements such as O and Cl (as well as other trace elements) become significant in their contribution to $f_{e}$ at low pressures $p_{\rm gas}\lesssim10^{-10}$~bar.

For molecular species the situation is more complicated (see lower panels, Figs.~\ref{gp_s_1}--\ref{md_m_1}).  In our analysis of oxygen-rich ultra-cool atmospheres of low mass objects, we have focused on several key molecules e.g.  H$_{2}$ for its high number density; H$_{2}$O for its spectral importance; and CO because it is little affected by the formation of dust clouds.  In general, as in the atomic case, the low-metallicity models (Figs.~\ref{gp_m_1}--\ref{md_m_1}) yield lower degrees of ionization for the participating molecules in comparison to the solar abundance cases (Figs.~\ref{gp_s_1}--\ref{md_s_1}).  In all the atmospheric models considered H$_{2}$, CO, H$_{2}$O, N$_{2}$ and SiO all yield $f_{e}\geq10^{-7}$ for all atmospheric pressures ($f_{e}\approx1,~10^{-4},~10^{-4},~10^{-5}$~and~$10^{-5}$~respectively) and so these species would successfully form a molecular plasma.  AlOH and Al$_{2}$O consistently yield $f_{e}\approx10^{-6}$ except at high pressures in low metal GP and BD models (Figs.~\ref{gp_m_1}~and~\ref{bd_m_1}) and in MD atmospheres (Figs.~\ref{md_s_1}~and~\ref{md_m_1}).  At high atmospheric pressures the local temperature is high enough to dissociate AlOH and Al$_{2}$O molecules, reducing their number density and hence the degree of ionization resulting from Alfv\'{e}n ionization.  Similarly, SiS produces modest degrees of ionization $f_{e}\approx10^{-6}$, except at low atmospheric pressures in low-metallicity MD model atmospheres (Fig.~\ref{md_m_1}) where $f_{e}<10^{-7}$. 
\FloatBarrier

\section{Discussion\label{section4}}

Observations of radio and X-ray emission suggest that such objects harbour an atmospheric, magnetized plasma.  This further suggests that substellar atmospheres are composed of gas-plasma mixtures that participate in a range of plasma phenomena such as waves, flows and instabilities.  In a plasma, dust becomes charged and can grow (forming clouds) electrostatically via ion deposition.  These charged dust particles will be susceptible to inter-grain electrical discharges that will energize the ambient gas-plasma and will participate in other collective plasma processes.  Therefore, it is key to our understanding of substellar atmospheres that the relevant plasma phenomena are incorporated into atmospheric models and their observational consequences quantified.  Moreover, it is critical to identify the ionization processes that create and replenish the atmospheric plasma.

The primary aim of this paper was to show that Alfv\'{e}n ionization is a significant process that can occur in the atmospheres of low-mass objects, responsible for the production of ionized pockets of gas that have a degree of ionization $\geq10^{-7}$, qualifying them as atmospheric plasmas.  A plasma is defined as a collection of charged particles, of sufficiently high number density that the coulomb force is significant in determining its properties, yet sufficiently dilute that the nearest neighbour interaction is dominated not by binary collisions, but instead by the collective electromagnetic influence of the many distant particles.  For an ionized gas to behave like a plasma rather than a neutral gas, the degree of ionization should be $\geq10^{-7}$.  The presence of pockets of atmospheric magnetized plasma opens the door to a multitude of plasma processes not yet considered in current atmospheric models, such as the growth of dust grains via ion deposition~\citep{stark2006}. 

This paper has outlined the criteria required for Alfv\'{e}n ionization to occur and has demonstrated it's application using example atmospheres of low-mass objects such as:  giant gas planets ($\log{g}=3.0$, $T_{\rm eff}=1500$~K); brown dwarfs ($\log{g}=5.0$, $T_{\rm eff}=1500$~K) and M-dwarfs ($\log{g}=4.0$, $T_{\rm eff}=2700$~K), for both  solar ([M/H]$=0.0$) and sub-solar ([M/H]$=-0.6$) metallicities.

Alfv\'{e}n ionization requires: (i) a low-density, magnetized seed plasma initially present in the atmosphere; and (ii) a neutral gas flow that attains a critical speed $v_{c}$.  Processes naturally occurring in the atmosphere such as thermal ionization, local discharge events (lightning), and cosmic-ray ionization, easily provide enough electrons and ions to constitute a seed plasma.  For the seed plasma to be magnetized an ambient magnetic field must be present and be sufficiently strong that the plasma dynamics are influenced by it.  For the model atmospheres considered here at high altitudes ($p_{\rm gas}\approx 10^{-12}$~bar) $B\gtrsim10^{-6}$~G is required locally, whereas at lower altitudes the criterion is much more severe $B\gtrsim10^{6}$~G; however, over a large atmosphere pressure range ($p_{\rm gas}\lesssim10^{-2}$~bar) the magnetized plasma criterion is easily achievable and the minimum magnetic flux density required locally is $\lesssim10^{4}$~G.  Typical average, global magnetic flux densities of gas planets, brown dwarfs and M-dwarfs ($\approx 10$~G for GPs and $1$~kG for BDs and MDs) infer that the seed plasma is adequately magnetized and that only at high pressures is a significant magnetic flux density required. Due to the weaker magnetic field strength of GPs, it is harder to ensure that the seed plasma is magnetized.  Therefore, Alfv\'{e}n ionization is easier to achieve in BDs and MDs.  However, other processes can occur in the exoplanetary environment that can enhance the ambient magnetic field (e.g.~\cite{khoda2012} discuss the formation of magnetodisks), aiding Alfv\'{e}n ionization.

For Alfv\'{e}n ionization the neutral gas flow speed must reach the critical flow speed $v_{c}$ which, for most gas species of interest $v_{c}\approx O(1-10$~kms$^{-1})$.  Calculations presented here considered three standard meteorological balanced flows (geostrophic, cyclostrophic and gradient) and found that reasonably low pressure gradients ranging from  $\approx10^{-18}-10^{-7}$~bar~cm$^{-1}$ are sufficient to drive the required neutral gas flows for Alfv\'{e}n ionization.  For the Earth, typical pressure gradients are of the order $|\partial p_{\rm gas}/\partial x|\approx10^{-7}$~bar~cm$^{-1}$ in the troposphere or  $\approx10^{-9}$~bar~cm$^{-1}$ at mid-latitudes, with much greater pressure gradients expected at fronts.  Bearing this in mind it seems plausible that in more exotic environments, such as the atmosphere of a low-mass object, such pressure gradients, and associated wind speeds, required for Alfv\'{e}n ionization are achievable.  Our analysis shows that at low atmospheric pressures the balanced flow criteria is smaller than at higher pressures, coinciding with the region of the atmosphere where the criteria for a magnetized seed plasma is most favourable.  In contrast to the other atmospheric models, GP and BD models for [M/H]$=0.0$ extend to lower mass densities and pressures and so in these atmospheres where the pressure is low, it is easier to fulfil the criteria for the balanced flows discussed here.

If the criteria for Alfv\'{e}n ionization can be met, then the process can ionize the entirety of the gas in a localized atmospheric volume, creating a plasma with an electron number density equal to the target neutral species number density plus the initial seed plasma number density.  In general,  if in a localized atmospheric pocket a particular species can be $100\%$ ionized, then the species with the greatest number density will yield the highest degree of ionization.  100\% ionization via the Alfv\'{e}n mechanism is perfectly feasible and has been demonstrated for Hydrogen in laboratory experiments~\citep{fahleson1961}.  Alfv\'{e}n ionization operates best at mid to high altitudes (mid to low atmosperhic pressures) where the seed plasma is easier to magnetize and the pressure gradients required to drive the required neutral flows are the smallest.  For the model atmospheres considered here, our results show that He, Fe, Mg and Na if ionized, increase the degree of ionization of the local environment beyond $10^{-7}$ ($f_{e}\approx1,~10^{-5},~10^{-5}$~and~$10^{-6}$~respectively) such that the ionized gas will be a plasma.  Furthermore, our calculations infer that it is easier to create a plasma with a higher degree of ionization in atmospheres with solar metallicities than with sub-solar metallicities.

Molecular plasmas are also expected to form in substellar atmospheres and can be easier to ionize in comparison to atoms since the required $v_{c}$ tends to be lower due to their larger mass and comparable ionization potential to their atomic constituents.  For the selected molecules studied here we find that Alfv\'{e}n ionization of H$_{2}$, CO, H$_{2}$O, N$_{2}$ and SiO all yield $f_{e}\geq10^{-7}$ for the model atmospheres considered here.  However, in some cases the ionization energy may exceed the dissociation energy of the molecule, increasing the likelihood that through the Alfv\'{e}n ionization mechanism certain molecules will be destroyed, lowering their local number density.  For example, CO has a dissociation energy of 11.13~eV (at 0 K) and an ionization potential of 14.0~eV,  implying that there is a chance that CO will be dissociated before it can be ionized.  For more complex molecules, total dissociation into neutral atoms will most likely occur in multiple stages e.g. H$_{2}$O$\rightarrow$OH+H and then OH$\rightarrow$O+H, with the energy required for each stage being 5.08~eV and 4.41~eV (at 0 K) respectively.  Furthermore, additional processes affecting the atmospheric chemistry can occur such as molecular dissociative recombination or dissociative attachment.

The degree of ionization as result of the Alfv\'{e}n ionization process easily outstrips that from thermal ionization, where $f_{\rm e,therm}\leq10^{-8}$ for most atmospheric pressures.  At high atmospheric pressures, where the temperature increases, $f_{\rm e,therm}$ will be enhanced, but never to levels equivalent to that obtained from Alfv\'{e}n ionization.  Cosmic-ray ionization can yield higher degrees of ionization ($f_{\rm e, cosmic}\approx10^{-8}$) relative to thermal considerations, but still cannot compete with the degrees of ionization from Alfv\'{e}n ionization.  In addition to Alfv\'{e}n ionization, inter-grain electrical discharges will enhance the electron number density, boosting the local degree of ionization~\citep{helling2011a}.

Alfv\'{e}n ionization also has important consequences for the subsequent chemistry that occurs in the atmospheres of low-mass objects due to the injection of a significant amount of electrons, positive ions and radicals.  This allows access to more complex chemistry, such as electron-moderated chemistry, that is otherwise not permitted if only thermal processes are available.  The presence of an amplified free electron population increases the likelihood that these electrons attach themselves to the ambient neutral atoms, molecules or dust particles enhancing the consequent chemistry including the formation of dust, or the formation of large-scale charge separations that may lead to lightning~\citep{helling2013}

Charged particle surfaces grow via the accumulation of neutral atoms and an electrostatically attracted ion flux (and also polarizable molecules, which will be attracted by the unbalanced charge too) and so will grow faster than the uncharged scenario when there is a neutral flux only.  This is assuming that the neutral flux remains the same and is not reduced by the creation of ions (at the expense of the neutrals) more than the charged grain's enhanced ion flux.  The excitation of metastable states can also affect the chemistry by momentarily locking energy in an excited state and therefore impacting on the energy budget of the system under consideration.  An observable consequence of a metastable population will be the presence of forbidden lines in observed spectra which would otherwise be absent in the non-ionising case.  For example, the metastable state 2$^{3}$S$_{1}$ ($E=19.82$~eV, $\nu\approx10^{15}$~Hz) of He, may produce a forbidden line in the spectra.

If low-mass object atmospheres harbour a magnetized plasma then it could be observable.  The motion and mutual interaction of the plasma particles would manifest itself as continuum Bremsstrahlung emission with a characteristic spectrum depending on the nature of the source.  For example,  the power radiated per unit volume per unit solid angle per unit angular frequency interval (the Bremsstrahlung emission coefficient, $\epsilon_{\omega}$) for a thermal plasma is given by
\begin{equation}
\epsilon_{\omega}(T_{e})=\frac{8}{3\sqrt{3}}\frac{Z^{2}n_{e}n_{i}}{m^{2}_{e}c^{3}}\left(\frac{e^{2}}{4\pi\epsilon}\right)^{3}\left(\frac{m_{e}}{2\pi k_{B}T_{e}}\right)^{1/2}\bar{g}(\omega,T_{e})\exp{\left(-\frac{\hbar\omega}{k_{B}T_{e}}\right)},
\end{equation}
where $\bar{g}$ is the Gaunt factor; $Z_{i}$ is the ion charge number; $\omega$ is the photon angular frequency; $n_{e}$ and $n_{i}$ is the electron and ion number density of the plasma respectively~\citep{boyd}.  In the regime $\hbar\omega\geq k_{B}T_{e}$ the slope of a log-linear plot of the Bremsstrahlung emission coefficient yields a measure of the electron temperature $T_{e}$.  Furthermore, if the electron temperature is known, the plasma density could be determined using the Bremsstrahlung emission coefficient~\citep{boyd}.

In addition to this, cyclotron/synchrotron emission would also be expected and would allow the ambient magnetic flux density to be diagnosed.  Superimposed on the continuum plasma emission would be emission and absorption lines from the 
plasma ion species (e.g. He, Mg, Fe, Na or H$_{2}$, CO, H$_{2}$O, N$_{2}$) constituting the atmospheric plasma, the identification of which would allow the characterisation of the plasma species.  Furthermore, the electron bombardment of the neutral species populating the atmosphere (as part of the Alfv\'{e}n ionization process) may excite the neutrals that then relax producing emission that could be identified in the spectra e.g. H$_{2}$, H$_{2}$O, CO, etc.  This may include rotational and vibrational excitation, resulting in discrete (and potentially time-dependent) emission lines.

The occurrence of Alfv\'{e}n ionization is not just restricted to the atmospheres of low-mass objects and may be applicable in other astrophysical environments.  In principle, Alfv\'{e}n ionization can occur in any astrophysical system where there is a magnetized seed plasma; and a neutral component that moves relative to the plasma exceeding the critical speed.  For example, protoplanetary disks~\citep{woitke2009} are composed of gas-plasma mixtures that have significant rotational flow speeds due to the gravitational accretion onto the central star.  Photoionization is thought to be the main contributor to the plasma; but Alfv\'{e}n ionization could significantly enhance the plasma density and help contribute to the generation of the stellar systems fossil magnetic field.  Alfv\'{e}n ionization could also be at work in the atmospheres of other stellar types such as a G-type star (e.g. the Sun), where flow speeds from emerging magnetic flux (i.e. a sun-, starspot) could exceed the critical speed resulting in ionization and enhancing the population of certain ions~\citep{diver2005}.

\acknowledgments

ChH, CRS and PBR are grateful for the financial support of the European Community under the FP7 by an ERC starting grant.  DAD is grateful for funding from the UK Science and Technology Funding Council via grant number ST/I001808/1.

\appendix

\section{Cosmic-ray ionization of substellar atmospheres\label{c_ray}}

Cosmic ray bombardment of a substellar atmosphere will increase $n_{e}$ via impact ionization of the initially neutral atmosphere.   Fig.~\ref{cos_ray} shows the resulting electron number density $n_{e}$ from the bombardment of a brown dwarf atmosphere (BD, [M/H] $=0.0$) by cosmic rays.  The plot shows that high in the atmosphere ($p_{\rm gas}\approx10^{-4}$~bar) cosmic ray ionization processes can increase the ambient electron number density to $n_{e}\approx10^{4}$~cm$^{-3}$.  In the calculation exhibited in Fig.~\ref{cos_ray}, the cosmic ray propagation is determined using a 1D Monte Carlo model involving 10000 cosmic rays that are each binned to different energies according to the spectrum proposed by ~\cite{nath1994}. Each cosmic ray is assigned a random number on a uniform distribution between zero and one. The cosmic rays then propagate through a column of atmospheric gas $\Delta N$. The random number assigned to the cosmic ray is compared to the cosmic-ray "optical depth" $\sigma(E)\Delta N$, where $\sigma(E)$ is the cross-section for an ionising collision between a proton and a hydrogen atom~\citep{padovani2009}. If the random number is $< \sigma(E)~\Delta N$, then the cosmic ray ionizes the hydrogen atom and loses a given amount of energy $\Delta W (E)$~\citep{rimmer2012}. The value of  $\Delta N$ is always chosen such that $\sigma(E)\Delta N < 1$. This process is repeated until the entire atmosphere is traversed. At each step, energy loss due to Alfv\'{e}n waves generated by the cosmic ray anisotropy is accounted as discussed in~\cite{skilling1976}. This is the only magnetic field effect considered.

\section{Atmospheric flows\label{appendix}}
Balanced flows are an idealised view of atmospheric motion, but give a good approximation of the dynamical behaviour and expected characteristic flow speeds.  In this appendix we consider a simple fluid dynamical model to describe three standard atmospheric balanced flows: geostrophic, cyclostrophic and gradient flows.  In the subsequent analysis we follow the texts of~\cite{vallis2008,holton2004,lavega2011,jacobson2005}.  Note that in the presence of significant ionization, atmospheres will be composed of a gas-plasma mixture and the dynamical evolution of the system will be affected by electromagnetic fields.  The onward evolution of the gas-plasma mixture would be much more complex since the two fluids (plasma and neutral gas) could be collisionally coupled~\citep{diver2006}.  In this scenario, the plasma could be described by magnetohydrodynamics (MHD,  for the governing equations see~\cite{boyd}), encapsulating the effect of the magnetic field on the charged species.  In the low-frequency, MHD regime the plasma can support a variety of plasma wave and flow phenomena (e.g.~\cite{diver2006,zaq2007,nak2005}).  Here, we consider an atmosphere predominantley composed of a neutral gas such that the dynamical effect of the plasma is a second order effect.

In a rotating reference frame the atmospheric gas dynamics are described by
\begin{eqnarray}
\frac{\partial \rho_{\rm gas}}{\partial t}+\nabla\cdot(\rho_{\rm gas} \mathbf{u})&=&0, \\
\rho_{\rm gas}\frac{\textnormal{D} \mathbf{u}}{\textnormal{D} t}&=&-\nabla p-2\rho_{\rm gas}\mathbf{\Omega}\times\mathbf{u}+\rho_{\rm gas}\mathbf{\hat{g}}, \\
\frac{\textnormal{D}p_{\rm gas}}{\textnormal{D}t}&=&-\gamma p\nabla\cdot\mathbf{u}+S, \\
p&=&h(T,\rho_{\rm gas}),
\end{eqnarray}
where
\begin{equation}
\mathbf{\hat{g}}=\mathbf{g}-\mathbf{\Omega}\times(\mathbf{\Omega}\times \mathbf{r})
\end{equation}
is the effective gravitational acceleration; $\gamma=5/3$ is the ratio of specific heats; and $h$ is the equation of state.  The momentum equation includes the pressure gradient force ($-\nabla p$),  the Coriolis effect ($-2\rho_{\rm gas}\mathbf{\Omega}\times\mathbf{u}$), the centrifugal effect ($-\rho_{\rm gas}\mathbf{\Omega}\times(\mathbf{\Omega}\times r)$) and gravity ($\rho_{\rm gas}\mathbf{g}$).  The energy equation includes an energy source and sink term $S$, describing the rate of heating or cooling of the gas.  The temptation is to cast these equations in spherical coordinates with a radial component defined by gravity $\mathbf{g}$.  However, the presence of a centrifugal effect complicates matters and so it is convenient to define the radial direction rather by $\mathbf{\hat{g}}$, encapsulating the centrifugal effect with gravity.  Strictly speaking the surfaces defined by the normal $\mathbf{\hat{g}}$ are not spherical but spheroidal, but for simplicity will be treated as spherical.   As a result, the horizontal component of effective gravity is zero thus simplifying the governing equations.  Additionally, further simplification can be sought by focussing on phenomena on a length scale where the sphericity of our rotating body is negligible.  In this case, instead of solving the equations in spherical coordinates we can use a local cartesian tangent plane on the surface of the sphere (this is known as the f-plane).

Consider a location with latitude $\phi$ defined on a surface with normal $\mathbf{\hat{g}}$ rotating around the north-south axis.  A local cartesian coordinate system can be set up with the x-axis horizontally due east, the y-axis horizontally due west and the z-axis vertically upwards.  The angular velocity can be written as $\mathbf{\Omega}=\omega(0,\cos{\phi},\sin{\phi})$ therefore, the governing dynamical equations become
\begin{eqnarray}
\frac{\textnormal{D}u_{x}}{\textnormal{D}t}&=&-\frac{1}{\rho_{\rm gas}}\frac{\partial p_{\rm gas}}{\partial x}+fu_{y}, \\
\frac{\textnormal{D}u_{y}}{\textnormal{D}t}&=&-\frac{1}{\rho_{\rm gas}}\frac{\partial p_{\rm gas}}{\partial y}-fu_{x}, \\
\frac{\textnormal{D}u_{z}}{\textnormal{D}t}&=&-\frac{1}{\rho_{\rm gas}}\frac{\partial p_{\rm gas}}{\partial z}+\hat{g},
\end{eqnarray}
where $f=2\omega\sin{\phi}$ is the Coriolis parameter characterising the significance of the Coriolis effect as a function of position on the surface.  If we assume that the vertical velocity component is negligible and the vertical component of the Coriolis effect is small in comparison to the effective gravity we can restrict our attention to a horizontal plane (i.e. $u_{z}=0$). This is the hydrostatic and traditional approximations respectively, yielding the so-called primitive equations of motion.  

\subsection{Geostrophic flow}
The Rossby number is defined as
\begin{equation}
Ro=\frac{|(\mathbf{u}\cdot\nabla)\mathbf{u}|}{|\mathbf{\Omega}\times\mathbf{u}|}=\frac{U}{Lf},
\end{equation}
characterising the significance of the Coriolis effect, hence rotation on the system.  $Ro\ll1$ defines a system strongly affected by the Coriolis effect and $Ro\gg1$ defines one where inertial effects dominate.  In the steady state ($\partial/\partial t =0$), if the local Rossby number is small ($Ro\ll1$) then a geostrophic wind balance is achieved
\begin{eqnarray}
u_{x}&=&-\frac{1}{f\rho_{\rm gas}}\frac{\partial p_{\rm gas}}{\partial y}, \\
u_{y}&=&\frac{1}{f\rho_{\rm gas}}\frac{\partial p_{\rm gas}}{\partial x}.
\end{eqnarray}
Geostrophic wind balance results from the balance between the Coriolis effect and the pressure gradient force.  Note that geostrophic flows are invalid at the equator because $f=0$.

\subsection{Cyclostrophic flow}
Transforming from cartesian to cylindrical polar coordinates the primitive equations of atmospheric motion in the f-plane become
\begin{eqnarray}
\rho_{\rm gas}\mathcal{D}u_{r}&=&-\frac{\partial p_{\rm gas}}{\partial r}+\rho_{\rm gas} f+\rho_{\rm gas}\frac{u_{\theta}^{2}}{r}, \label{mr} \\
\rho_{\rm gas}\mathcal{D}u_{\theta}&=&-\frac{1}{r}\frac{\partial p_{\rm gas}}{\partial \theta}-\rho_{\rm gas} f-\rho_{\rm gas}\frac{u_{\theta}u_{r}}{r}, \label{mt} \\
\rho_{\rm gas}\mathcal{D}u_{z}&=&-\frac{\partial p_{\rm gas}}{\partial z}+\rho_{\rm gas}\hat{g}, \\
\end{eqnarray}
where
\begin{equation}
\mathcal{D}=\frac{\partial}{\partial t}+u_{r}\frac{\partial}{\partial r}+\frac{u_{\theta}}{r}\frac{\partial}{\partial \theta}+u_{z}\frac{\partial}{\partial z}.
\end{equation}
The latter terms in the $r$ and $\theta$ components of the momentum equation (Eq.~\ref{mr}~and~\ref{mt}, respectively) correspond to the local centrifugal ($u_{\theta}^{2}/r$) and local Coriolis effects ($u_{\theta}u_{r}/r$), as experienced in the non-inertial frame of a rotating fluid parcel of atmospheric gas.  Cyclostrophic flows are characterised by the balance between the pressure gradient force and the centrifugal effect.  This is applicable when the horizontal length scale under consideration is small enough that the global Coriolis effect ($\rho_{\rm gas} f$) can be neglected.    In the steady state, assuming the hydrostatic approximation, uniform circular motion ($u_{r}=0$, $\partial p_{\rm gas}/\partial\theta=0$), where the centripetal force is provided by the pressure gradient force, the cyclostrophic approximation is
\begin{equation}
\frac{\rho_{\rm gas} u_{\theta}}{r}=\frac{\partial p_{\rm gas}}{\partial r},
\end{equation}
yielding a cyclostrophic wind speed,
\begin{equation}
u_{\theta}=\left(\frac{r}{\rho_{\rm gas}}\frac{\partial p_{\rm gas}}{\partial r}\right)^{1/2}.
\end{equation}

\subsection{Gradient flow}
Retaining the primitive equations in a local cylindrical geometry, another force balance can be obtained between the global Coriolis effect, the pressure gradient force and the local centrifugal effect.  For simplicity, uniform circular motion is assumed ($u_{r}=0$), the hydrostatic approximation is assumed and $\partial p_{\rm gas}/\partial\theta=0$. This yields
\begin{equation}
-\frac{\rho_{\rm gas} u^{2}_{\theta}}{r}=-\frac{\partial p_{\rm gas}}{\partial r}+\rho_{\rm gas} f u_{\theta}.
\end{equation}
Solving the quadratic, the gradient flow speed is
\begin{equation}
u_{\theta}=-\frac{1}{2}rf\pm\frac{1}{2}r\left(f^{2}+\frac{4}{r\rho_{\rm gas}}\frac{\partial p_{\rm gas}}{\partial r} \right)^{1/2}.
\end{equation}

\clearpage


\begin{thebibliography}{}

\bibitem[Alfv\'{e}n(1960)]{alfven1960} Alfv\'{e}n, H. 1960, Rev. Mod. Phys., 32, 4, 710

\bibitem[Antonova {\it et al.}(2007)]{antonova2007} Antonova, A., Doyle, J. G., Hallinan, G., Golden, A., Koen, C. 2007, A\&A, 472, 257

\bibitem[Antonova {\it et al.}(2008)]{antonova2008} Antonova, A., Doyle, J. G., Hallinan, G., Bourke, S., Golden, A. 2008, A\&A, 487, 317

\bibitem[Antonova {\it et al.}(2013)]{antonova2013} Antonova, A., Hallinan, G., Doyle, J. G., Yu, S., Kuznetsov, A., Metodieva, Y., Golden, A., Cruz, K. L. 2013, A\&A, 549, A131

\bibitem[Beebe(1997)]{beebe1997} Beebe, R. 1997, Jupiter the Giant Planet, 2nd Edition, (Washington, DC: Smithsonina Institution Press)

\bibitem[Becker {\it et al.}(2006)]{becker2006} Becker, K. H., Schoenbach, K. H., Eden, J. G. 2006, J. Phys. D: Appl. Phys., 39, R55

\bibitem[Berger(2006)]{berger2006} Berger, E. 2006, \apj, 648, 629

\bibitem[Berger {\it et al.}(2009)]{berger2009} Berger, E., Rutledge, R. E., Phan-Bao, N., Basri, G., Giampapa, M. S., Gizis, J. E., Liebert, J., Mart\'{i}n, E., Fleming, T. A. 2009, \apj, 695, 310

\bibitem[Borrero {\it et al.}(2008)]{borrero2008} Borrero, J. M., Lites, B. W., Solanki, S. K. 2008, A\&A, 481, L13

\bibitem[Boyd \& Sanderson(2003)]{boyd} Boyd, T. J. M., \& Sanderson J. J. 2003, The Physics of Plasmas, (Cambridge, UK:  Cambridge University Press) 

\bibitem[Bransden \& Joachin(2000))]{bransden2000} Bransden, B. H., \& Joachin, C. J. 2000, Quantum Mechanics, 2nd Edition, (Addison-Wesley)

\bibitem[Brenning(1992)]{brenning1992} Brenning, N. 1992, Space Sci. Rev., 59, 209

\bibitem[Chang {\it et al.}(2010)]{chang2010}  Chang, Z. S., Zhao, N., Yuan, P. 2010, Phys. Plasmas, 17, 113514

\bibitem[Choi {\it et al.}(2010)]{choi2010} Choi, D. S., Showman, A. P. Vasavada, A. R. 2012, Icarus, 207, 359

\bibitem[Christensen {\it et al.}(2009)]{christensen2009} Christensen, U. R., Holzwarth, V., Reiners, A. 2009, Nature, 457, 168

\bibitem[Cooper \& Showman(2005)]{cooper2005} Cooper, C. S., \& Showman, A. P. 2005, ApJ, 629, L45

\bibitem[Danielsson(1973)]{danielsson1973} Danielsson, L. 1973, Astrophys. Space Sci., 24, 459

\bibitem[Dehn(2007)]{dehn2007} Dehn, M. 2007, PhD thesis, Univ. Hamburg

\bibitem[Diver {\it et al.}(2005)]{diver2005} Diver, D. A., Fletcher, L., Potts, H. E. 2005,  Sol. Phys., 227, 207

\bibitem[Diver {\it et al.}(2006)]{diver2006} Diver, D. A., Potts, H. E., Teodoro, L. F. A. 2006, NJoP, 8, 265

\bibitem[Diver(2013)]{diver2001} Diver, D. A. 2013, A plasma formulary for physics, technology, and astrophysics, 2nd Edition, (Berlin; New York: Wiley-VCH)

\bibitem[Dobbs-Dixon \& Lin(2008)]{dobbs2008} Dobbs-Dixon, I., \& Lin, D. N. C. 2008, ApJ, 673, 513

\bibitem[Dobbs-Dixon {\it et al.}(2010)]{dobbs2010} Dobbs-Dixon, I., Cumming, A., Lin, D. N. C. 2010, ApJ, 710, 1395

\bibitem[Dobbs-Dxion {\it et al.}(2012)]{dixon2012} Dobbs-Dixon, I., Agol, E., Burrows, A. 2012, \apj, 751, 87

\bibitem[Donati \& Landstreet(2009)]{donati2009} Donati, J.-F., \&Landstreet, J. D. 2009, \araa, 47, 333

\bibitem[Dunning \& Hulet(1996)]{dunning1996} Dunning F. B., Hulet, R. G. 1996, Atomic, Molecular, and Optical Physics: Atoms and Molecules: Volume 29b: Atomic, Molecular, and Optical Physics, (London:  Academic Press)

\bibitem[Fahleson(1961)]{fahleson1961} Fahleson, U. V. 1961, Phys. Fluids, 4 123

\bibitem[Fridman(2008)]{fridman2008} Fridman, A. 2008, Plasma Chemistry, 1st Edition, (New York:  Cambridge University Press)

\bibitem[Guo {\it et al.}(2009)]{guo2009} Guo, Y., Yuan, P., Shen, X., Wang, J. 2009, Phys. Scr., 80, 035901

\bibitem[Hallinan {\it et al.}(2006)]{hallinan2006} Hallinan, G., Antonova, A., Doyle, J. G., Bourke, S., Brisken, W., Golden, A.  2006, \apj, 653, 690

\bibitem[Hallinan {\it et al.}(2007)]{hallinan2007} Hallinan, G., Bourke, S., Lane, C., Antonova, A., Zavala, R. T., Brisken, W. F., Boyle, R. P., Vrba, F. J., Doyle, J. G., Golden, A.  2007, \apj, 663, L25-L28

\bibitem[Hallinan {\it et al.}(2008)]{hallinan2008} Hallinan, G., Antonova, A., Doyle, J. G., Bourke, S., Lane, C., Golden, A.  2008, \apj, 684, 644

\bibitem[Hammel {\it et al.}(2001)]{hammel2001} Hammel, H. B., Rages, K., Lockwood, G. W., Karkoschka, E., de Peter, I.  2001, Icarus, 153, 229

\bibitem[Hartman {\it et al.}(2005)]{hartman2005} Hartman, H., Schef, P., Lundin, P., Ellmann, A., Johansson, S., Lundberg, H., Mannervik, S., Norlin, L. -O., Rostohar, D., Royen, P. 2005, Mon. Not. R. Astron. Soc., 361, 206

\bibitem[Helling {\it et al.}(2008)]{helling2008} Helling, Ch., Dehn, M., Woitke, P., Hauschildt, P. H. 2008, ApJ, 675, L105

\bibitem[Helling {\it et al.}(2008b)]{helling2008b} Helling, Ch., Ackerman, A., Allard, F., Dehn, M., Hauschildt, P., Homeier, D., Lodders, K., Marley, M., Rietmeijer, F., Tsuji, T., Woitke, P.  2008b, MNRAS, 391, 1854

\bibitem[Helling {\it et al.}(2008c)]{helling2008c} Helling, Ch., Woitke, P., Thi, W.-F. 2008c, A\&A, 485, 547

\bibitem[Helling {\it et al.}(2011a)]{helling2011a} Helling, Ch., Jardine, M., Witte, S., Diver, D. A.  2011a, \apj, 727, 4

\bibitem[Helling {\it et al.}(2011b)]{helling2011b} Helling, Ch., Jardine, M., Mokler, F. 2011b, \apj, 737, 38

\bibitem[Helling {\it et al.}(2013)]{helling2013} Helling Ch., Jardine, M., Stark, C. R., Diver, D. A. 2013, \apj, 767, 136.

\bibitem[Heng {\it et al.}(2011)]{heng2011} Heng, K., Menou, K., Phillips, P. J. 2011, MNRAS, 413, 2380 

\bibitem[Heng(2012)]{heng2012} Heng, K. 2012, \apj, 761, L1

\bibitem[Holton(2004)]{holton2004} Holton, J. R.  2004, An Introduction to Dynamic Meteorology, 4th Edition, (Burlington; San Diego; London:  Elsevier Academic Press)

\bibitem[Jacobson(2005)]{jacobson2005} Jacobson, M. Z. 2005, Fundamentals of Atmospheric Modeling, 2nd Edition, (Cambridge: Cambridge University Press)

\bibitem[Joergens {\it et al.}(2003)]{joergens2003} Joergens, V., Fern\'{a}ndez, M., Carpenter, J.M., Neuh\
"{a}user, R. 2003, \apj, 594, 971

\bibitem[Khodachenko {\it et al.}(2012)]{khoda2012} Khodachenko, M. L., Alexeev, I., Belenkaya, E., Lammer, H., Grei{\ss}meier, J.-M., Leitzinger, M., Odert, P., Zaqarashvili, T., Rucker, H. O. 2012, \apj, 744, 70

\bibitem[Lai(2001)]{lai2001} Lai, S. T. 2001, Rev. Geophys., 39, 4, 471

\bibitem[Lewis {\it et al.}(2010)]{lewis2010} Lewis, N. K., Showman, A. P., Fortney, J. J., Marley, M. S., Freedman, R. S. 2010, ApJ, 720, 344

\bibitem[Li {\it et al.}(2009)]{li2009} Li, C., Ebert, U., Hundsdorfer, W. 2009, J. Phys. D:  Appl. Phys., 42, 202003

\bibitem[Machida \& Goertz(1986)]{machida1986} Machida, S., \& Goertz, C. K. 1986, J. Geophys. Res. Space Physics, 91, 11965

\bibitem[MacLachlan {\it et al.}(2009)]{craig2009} MacLachlan, C. S., Diver, D. A., Potts, H. E. 2009, New J. Phys., 11, 063001

\bibitem[MacLachlan {\it et al.}(2013)]{craig2013} MacLachlan, C. S., Potts, H. E., Diver, D. A.  2013, Plasma Sources Sci. Technol., 22, 015025

\bibitem[McBride {\it et al.}(1972)]{mcbride1972} McBride, J. B., Ott, E., Boris, J. P., Orens, J. H. 1972, Phys. Fluids, 15, 2367

\bibitem[McLean {\it et al.}(2011)]{mclean2011} McLean, M., Berger, E., Irwin, J., Forbrich, J., Reiners, A. 2011, \apj, 741, 27

\bibitem[McLean {\it et al.}(2012)]{mclean2012} McLean, M., Berger, E., Reiners, A. 2012, \apj, 746, 23

\bibitem[McNeil {\it at al.}(1990)]{mcneil1990} McNeil, W. J., Lai, S. T., Murad, E. 1990,  J. Geophys. Res. Space Physics, 95, 10345

\bibitem[Menou \& Rauscher(2009)]{menou2009} Menou, K., \& Rauscher, E., 2009, ApJ, 700, 887

\bibitem[Nakariakov \& Verwichte(2005)]{nak2005} Nakariakov, V. M., Verwichte, E. 2005, Living Rev.  Sol. Phys., 2, 3

\bibitem[Nath \& Biermann(1994)]{nath1994} Nath, B. B., \& Biermann, P. L. 1994, \mnras, 267, 447

\bibitem[Newell(1985)]{newell1985} Newell, P. T. 1985, Rev. Geophys., 123

\bibitem[Nichols {\it et al.}(2012)]{nichols2012} Nichols, J. D., Nurleigh, M. R., Caswell, S. L., Cowley, S. W. H., Wynn, G. A., Clarke, J. T., West, A. A. 2012, \apj, 760. 59

\bibitem[Osten {\it et al.}(2009)]{osten2009} Osten R. A., Phan-Bao, N., Hawley, S. L., Neill Reid, I., Ojha, R. 2009, \apj, 700, 1750

\bibitem[Padovani {\it et al.}(2009)]{padovani2009} Padovani, M., Galli, D., Glassgold, A. E. 2009, \aap, 501, 619

\bibitem[Person {\it et al.}(1990)]{person1990} Person, J. C., Resendes, D., Petschek, H., Hastings, D. E. 1990,  J. Geophys. Res. Space Physics, 95, 4039

\bibitem[Porco {\it et al.}(2005)]{porco2005} Porco, C. C., {\it et al.}  2005, Science, 307, 1243

\bibitem[Priest(1985)]{priest1985} Priest, E. R. (ed.) 1985, Solar System Magnetic Fields, 1st Edition,  (Dordrecht, Netherlands: D. Reidel Publ. Co.)

\bibitem[Rauscher \& Menou(2010)]{rauscher2010} Rauscher, E., \& Menou, K. 2010, ApJ, 714, 1334

\bibitem[Reiners(2012)]{reiners2012} Reiners, A. 2012,  Living Rev. Solar Phys., 8, 1

\bibitem[Reiners {\it et al.}(2012b)]{reiners2012b} Reiners, A., Joshi, N., Goldman, B. 2012b, \apj, 143, 93

\bibitem[Rimmer {\it et al.}(2012)]{rimmer2012} Rimmer, P. B., Herbst, E., Morata, E., Roueff, E. 2012, \aap, 537, A7

\bibitem[Rimmer {\it et al.}(2013)]{rimmer2013} Rimmer, P. B., Helling, Ch. 2013, Submitted to \apj, under review.

\bibitem[Route \& Wolszczan(2012)]{route2012} Route, M., \& Wolszczan, A. 2012, \apj, 747, L22

\bibitem[S\'{a}nchez-Lavega(2004)]{sanchez2004} S\'{a}nchez-Lavega, A. 2004, \apj, 609, L87

\bibitem[S\'{a}nchez-Lavega(2001)]{sanchez2001} S\'{a}nchez-Lavega, A. 2001, A\&A, 377, 360

\bibitem[S\'{a}nchez-Lavega(2011)]{lavega2011} S\'{a}nchez-Lavega, A. 2011, An Introduction to Planetary Atmospheres, 1st Edition, (Boca Raton: Taylor and Francis)

\bibitem[Scholz \& Eisl\"{o}ffel(2005)]{scholz2005} Scholz, A., \& Eisl\"{o}ffel, J. 2005, A\&A. 429, 1007

\bibitem[Showman \& Guillot(2002)]{showman2002} Showman, A. P., \& Guilot, T. 2002, A\&A, 385, 166

\bibitem[Showman {\it et al.}(2008)]{showman2008} Showman, A. P., Cooper, C. S., Fortney, J. J.,  Marley, M. S. 2008, ApJ, 682, 559

\bibitem[Showman {\it et al.}(2009)]{showman2009} Showman, A. P., Fortney, J. J.,  Lian, Y., Marley, M. S., Freedman, R. S., Knutson, H. A., Charbonneau, D. 2009, ApJ, 699, 564

\bibitem[Shulyak {\it et al.}(2011)]{shulyak2011} Shulyak, D., Seifahrt, A., Reiners, A., Kochukhov, O., Piskunov, N. 2011, \mnras, 418, 2548 

\bibitem[Skilling \& Strong(1976)]{skilling1976} Skilling, J., \& Strong, A. W. 1976, \aap, 53, 253

\bibitem[Sromovsky \& Fry(2005)]{sromovsky2005} Sromovsky, L. A., \& Fry, P. M. 2005, Icarus, 179, 459

\bibitem[Stark {\it et al.}(2006)]{stark2006} Stark, C. R., Potts, H. E., \& Diver, D. A. 2006, A\&A, 457, 365

\bibitem[Uman \& Orville(1964)]{uman1964} Uman, M. A., \& Orville, R. E. 1964, J. Geophys. Res., 69, 24, 5151

\bibitem[Vallis(2008)]{vallis2008} Vallis, G. V. 2008, Atmospheric and Oceanic Fluid Dynamics Fundamentals and Large-Scale Circulation, 3rd Edition, (Cambridge University Press)

\bibitem[Vidotto {\it et al.}(2011)]{vidotto2011} Vidotto, A. A., Jardine, M., Helling, Ch. 2011, Mon. Not. R. Astron. Soc., 411, L46

\bibitem[Vidotto {\it et al.}(2011b)]{vidotto2011b} Vidotto, A. A., Jardine, M., Helling, Ch. 2011, Mon. Not. R. Astron. Soc., 414, 1573

\bibitem[Witte {\it et al.}(2011)]{witte2011} Witte, S., Helling, Ch., Barman, T., Heidrich, N., Hauschildt, P. H. 2011, A\&A, 529, A44

\bibitem[Woitke \& Helling(2003)]{woitke2003} Woitke, P., Helling, Ch. 2003, A\&A, 399, 297
 
 \bibitem[Woitke {\it et al.}(2009)]{woitke2009} Woitke, P., Kamp, I., Thi, W.-F. 2009, A\&A, 501, 383
 
\bibitem[Yu {\it et al.}(2012)]{yu2012} Yu, S., Doyle, J. G., Kuznetsov, A., Hallinan, G., Antonova, A., MacKinnon, A. L., Golden, A. 2012, \apj, 752, 60 

\bibitem[Zapatero Osorio {\it et al.} (2006)]{osorio2006} Zapatero Osorio, M. R., Mart\'{i}n, E. L., Bouy, H., Tata, R., Deshpande, R. 2006, \apj, 647, 1405

\bibitem[Zaqarashvili {\it et al.}(2007)]{zaq2007} Zaqarashvili, T. V., Oliver, R., Ballester, J. L., Shergelashvili, B. M. 2007, A\&A, 470, 815

\bibitem[Zethson {\it et al.}(2001)]{zethson2001} Zethson, T., Gull, T. R., Hartman, H., Johansson, S., Davidson, K., Ishibashi, K.  2001, \apj, 122, 322



\end{thebibliography}
\end{document}